\documentclass[useAMS,usenatbib]{mn2e}

\usepackage{amsmath}
\usepackage{graphicx}
\usepackage{rotating}

\def\lesssim{\mathrel{\hbox{\rlap{\hbox{\lower4pt\hbox{$\sim$}}}\hbox{$<$}}}}
\def\gtrsim{\mathrel{\hbox{\rlap{\hbox{\lower4pt\hbox{$\sim$}}}\hbox{$>$}}}}

\def\arcmin{\hbox{$^\prime$}}
\def\arcsec{\hbox{$^{\prime\prime}$}}

\def\eg{{\it e.g.\ }}
\def\ie{{\it i.e.\ }}
\def\ltsima{$\; \buildrel < \over \sim \;$}
\def\simlt{\lower.5ex\hbox{\ltsima}}

\usepackage{xspace}
\usepackage[usenames]{color}
\usepackage{ulem}

\newcommand{\hb}[0]{H$\alpha$\xspace}
\newcommand{\rb}[0]{$r'$\xspace}
\newcommand{\ib}[0]{$i'$\xspace}
\newcommand{\ag}[0]{AstroGrid\xspace}

\begin{document}

\title[IPHAS Initial Data Release]{Initial Data Release from the INT Photometric H$\alpha$ Survey of the Northern Galactic Plane (IPHAS)}
\author[Gonz\'alez-Solares et al.]{E. A. Gonz\'alez-Solares$^1$\footnote{eglez@ast.cam.ac.uk},
N. A. Walton$^1$, R. Greimel$^{2,3}$, J. E. Drew$^{4,5}$, M. J. Irwin$^1$,\newauthor
 S. E. Sale$^4$, K. Andrews$^6$, A. Aungwerojwit$^{7,8}$, M. J. Barlow$^9$, E. van den Besselaar$^{10}$,
\newauthor
R. L. M. Corradi$^{2,11}$, B. T. G\"ansicke$^7$, P. J. Groot$^{10}$, A. S. Hales$^{9,12}$,  E. C. Hopewell$^4$,   \newauthor
H. Hu$^{10}$, J. Irwin$^{1,13}$, C. Knigge$^{14}$, E. Lagadec$^{15}$, P. Leisy$^{2,11}$, J. R. Lewis$^1$,
   \newauthor 
A. Mampaso$^{11}$,  M. Matsuura$^{15}$, B. Moont$^4$, L. Morales-Rueda$^{10}$, R. A. H. Morris$^{16}$,
 \newauthor
T. Naylor$^{17}$, Q. A. Parker$^{18,19}$, P. Prema$^1$, S. Pyrzas$^7$, G. T. Rixon$^1$, P. Rodr\'{\i}guez-Gil$^{7,11}$, \newauthor 
G. Roelofs$^{10,13}$, L. Sabin$^{15}$, I. Skillen$^2$, J. Suso$^{20}$, R. Tata$^{21}$, K. Viironen$^{11}$,
J. S. Vink$^{22}$,  
\newauthor
A. Witham$^{14}$,  N. J. Wright$^9$,   A. A. Zijlstra$^{15}$, 
A. Zurita$^{23}$, J. Drake$^{13}$, J. Fabregat$^{20}$,  
\newauthor
D. J. Lennon$^{2,11}$, P. W. Lucas$^{5}$, E. L. Mart\'{\i}n$^{11,21}$,  S. Phillipps$^{16}$, D. Steeghs$^{7,13}$, \newauthor Y. C. Unruh$^4$ \vspace{0.5em} \\
$^1$ Institute of Astronomy, Madingley Road, Cambridge CB3~0HA, U.K. \\
$^2$ Isaac Newton Group of Telescopes, Apartado de correos 321, E38700 Santa Cruz de La Palma, Tenerife, Spain \\
$^3$ Institut f\"ur Physik, Karl-Franzen Universit\"at Graz, Universit\"atsplatz 5, 8010 Graz, Austria \\
$^4$ Imperial College London, Blackett Laboratory, Exhibition Road, London, SW7 2AZ \\
$^5$ Centre for Astrophysics Research, University of Hertfordshire, College Lane, Hatfield AL10 9AB \\
$^6$ Royal Observatory Edinburgh, Blackford Hill, Edinburgh, EH9 3HJ, U.K.\\
$^7$ Department of Physics, University of Warwick, Coventry, CV4 7AL, U.K. \\
$^8$ Department of Physics, Faculty of Science, Naresuan University, Phitsanulok, 65000, Thailand \\
$^9$ University College London, Department of Physics \& Astronomy, Gower Street, London, WC1E 6BT, U.K. \\
$^{10}$ Department of Astrophysics/IMAPP, Radboud University Nijmegen, P.O. Box 9010, 6500 GL, Nijmegen, The Netherlands \\
$^{11}$ Instituto de Astrof\'{\i}sica de Canarias, Via L\'actea s/n, E38200 La Laguna, Santa Cruz de Tenerife, Spain \\
$^{12}$ National Radio Astronomy Observatory, 520 Edgemont Road, Charlottesville, Virginia, 22903-2475, USA \\
$^{13}$ Harvard-Smithsonian Center for Astrophysics, 60 Garden Street, Cambridge, MA 02138, U.S.A. \\
$^{14}$ School of Physics \& Astronomy, University of Southampton, Southampton, SO17 1BJ, U.K. \\
$^{15}$ School of Physics and Astronomy, University of Manchester, Sackville Street, PO Box 88, Manchester, M60 1QD, U.K. \\
$^{16}$ Astrophysics Group, Department of Physics, Bristol University, Tyndall Avenue, Bristol, BS8 1TL, U.K. \\
$^{17}$ School of Physics, University of Exeter, Stocker Road, Exeter, EX4 4QL, U.K. \\
$^{18}$ Department of Physics, Macquarie University, NSW 2109, Australia \\
$^{19}$ Anglo-Australian Observatory, PO Box 296, Epping NSW 1710, Australia \\
$^{20}$ Observatorio Astron\'omico de Valencia, Universidad de Valencia, 46071 Paterna-Valencia, Spain \\
$^{21}$ Physics Department, University of Central Florida, Orlando, FL 32816, U.S.A \\
$^{22}$ Armagh Observatory, College Hill, Armagh BT61 9DG, Northern Ireland \\
$^{23}$ Departamento de F\'{\i}sica Teorica y del Cosmos, Facultad de Ciencias, Av. Fuentenueva s/n, E18071 Granada, Spain 
}

\date{Received .................; Accepted ................}

\pagerange{\pageref{firstpage}--\pageref{lastpage}} \pubyear{2008}

\maketitle

\label{firstpage}

\begin{abstract}
The INT/WFC Photometric \hb Survey of the Northern Galactic Plane (IPHAS) is an imaging 
survey being carried out in \hb, \rb and \ib filters, with the Wide Field 
Camera (WFC) on the 2.5-metre Isaac Newton Telescope (INT) to a depth 
of \rb=20 (10$\sigma$).
The survey is aimed at revealing the large scale organisation of
the Milky Way and can be applied to identifying a range of stellar
populations within it. Mapping emission line objects enables a particular focus on
objects in the young and old stages of stellar evolution ranging
from early T-Tauri stars to late planetary nebulae.
In this paper we present the IPHAS Initial Data Release, primarily a
photometric catalogue of about 200 million unique objects, coupled
with associated image data covering about 1,600 square degrees in
three passbands. We note how access to the primary data products has
been implemented through use of standard virtual observatory
publishing interfaces. Simple traditional web access is provided to the main IPHAS
photometric catalogue, in addition to a number of common catalogues
(such as 2MASS) which are of immediate relevance.
Access through the AstroGrid VO Desktop opens up the full range
of analysis options, and allows full integration with the wider range
of data and services available through the Virtual Observatory.
The IDR represents the largest dataset published primarily through VO
interfaces to date, and so stands as an examplar of the future of survey
data mining.  Examples of data access are given, including a
cross-matching of IPHAS photometry with sources in the UKIDSS Galactic
Plane Survey that validates the existing calibration of the best data.

\end{abstract}


\section{Introduction}

The INT Photometric \hb Survey of the Northern Galactic Plane (IPHAS) is a 1800 deg$^2$ CCD survey of the northern Milky Way ($|b|<5^\circ$) using the \rb, \ib and \hb passbands (table~\ref{tab:filters}) down to \rb$=20$ (Vega, 10$\sigma$ for a point-like source in an aperture of 1.2 arcsec). The execution of IPHAS, the properties of its characteristic
colour-colour plane, and the likely scientific scope of the survey
were discussed by \cite{2005MNRAS.362..753D}

This paper describes the Initial Data Release (IDR) of the IPHAS survey, containing observations for about 200 million objects and comprising 2.4 Tb of processed imaging data. The survey is being carried out using the Wide Field Camera (WFC) on the INT telescope. The WFC consists of four 2048x4096 pixel CCDs with a pixel scale of 0.33 arcsec and a field of view of 0.29 square degrees. The camera CCD detectors are configured in an L-shape
and CCD number 3 is slightly vignetted in one corner.

The data have already been used to carry out a series of investigations: the interaction between the planetary nebula Sh2-188 and the inter-stellar medium \citep{2006MNRAS.366..387W}; the discovery of a new quadrupolar nebula \citep{2006A&A...458..203M};  the study of the properties of a sample of cataclysmic variables \citep{2006MNRAS.369..581W}; the interaction of the jet of Cygnus X-1 with the surrounding medium \citep{2007MNRAS.376.1341R}; brown dwarf searches (Mart\'{\i}n et al. 2008, submitted); detection of nova progenitors \citep{2007ATel.1031....1S,2006ATel..795....1S}; the study of the properties of symbiotic stars and first discoveries \citep{2007arXiv0712.2391C}, the study of young pre-main sequence stars in CygOB2 (Vink et al. 2008, submitted), the presentation of a sample of about 5000 \hb\ emitting stars \citep{2008MNRAS.tmp..127W} and the description of a method for selecting A stars and exploiting
them to estimate distances and reddening \citep{2008arXiv0802.3868D}. Several other investigations are  
on-going and we hope the data release described here will lead to many data 
mining programs by the community.

In addition to the optical data provided by IPHAS, the Galactic Plane Survey \citep[GPS;][]{2007arXiv0712.0100L} of the UKIDSS \citep{2006MNRAS.372.1227D} is also observing a similar area in J, H and K. 
This will be followed in the next few years by a survey of the southern Galactic Plane, VPHAS+, on the VLT Survey Telescope (VST) using OmegaCam~\citep{2004SPIE.5492..484K}. It serves as a digital successor
to the UK Schmidt H$\alpha$ Survey~\citep{2005MNRAS.362..689P} and will
reach about 2 magnitudes deeper. The UVEX project is targeting the same area in the U and g bands. The \textit{Herschel} observatory \citep{2005ESASP.577....3P} will also target the Galactic Plane. 
Together, these surveys will provide a better understanding of the stellar content
and properties of our Galaxy.

The aim of this paper is to provide a guide to the Initial Data Release of the IPHAS survey, describe the data products and the data properties and usage. Section~\ref{sec:summary} summarises the observations, sky coverage and data quality. The data processing is briefly described in section~\ref{sec:processing}. The data products contained in this release are described in section~\ref{sec:dataproducts}. Section~\ref{sec:science} shows an example usage of the catalogue to extract sources detected in a large area of sky and cross match with the GPS survey. Finally in section~\ref{sec:access} we describe the different user access interfaces to the data.

\begin{table}
\centering
\begin{tabular}{lrrr} \hline\hline
\textbf{Filter} & \textbf{Cen Wave (\AA)} & \textbf{Width (\AA)} & \textbf{AB (mag)} \\ \hline
$r'$ & 6254.1 & 1350.0 & 0.164 \\
H$\alpha$ & 6568.3 & 96.0 & 0.352 \\
$i'$ & 7771.9 & 1490.0 & 0.413 \\ \hline
\end{tabular}
\caption{Filter parameters. The second column lists the effective wavelength of the filter and the third its
width. The last column lists the value to be added to the Vega magnitude to convert it to AB. Filter profiles are available from the ING web site (www.ing.iac.es) and discussed in Drew et al. 2005.}
\label{tab:filters}
\end{table}

\section{Sky Coverage and Data Quality}
\label{sec:summary}

\begin{figure*}
\centering
\includegraphics[trim=10 20 10 10,clip=true,width=\hsize]{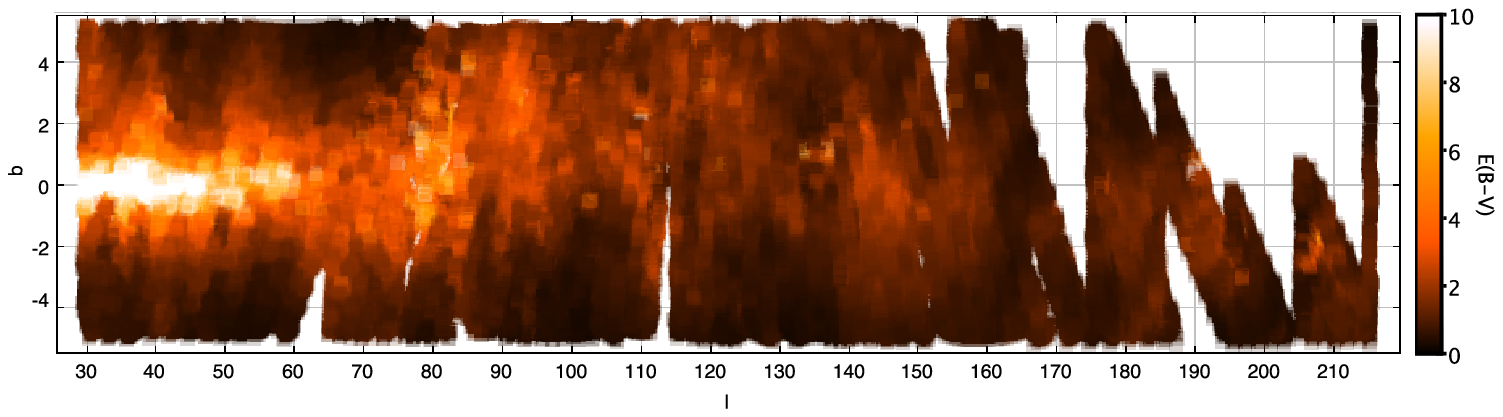}\\
\includegraphics[trim=10 20 10 10,clip=true,width=\hsize]{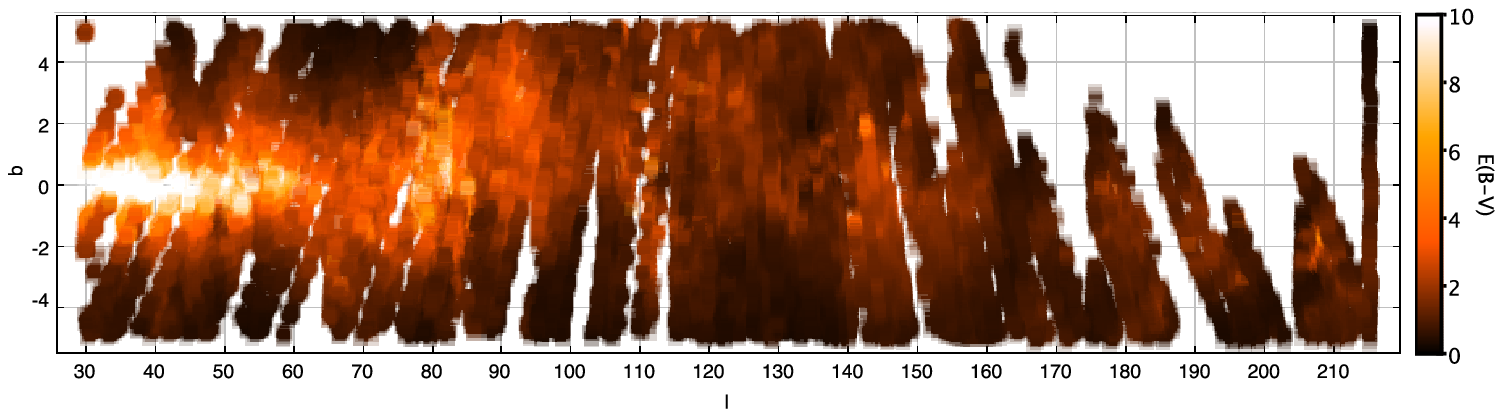}
\caption{IPHAS Initial Data Release sky coverage in galactic coordinates (degrees). The top figure displays the coverage of the full release and the bottom one the coverage of the \textbf{PhotoObjBest} best observations subset. Brightness indicates the total line-of-sight reddening as computed from the Schlegel et al. (1998) dust maps using 4 pixel interpolation.}
\label{fig:coverage}
\end{figure*}

The IDR contains 1,600 deg$^2$ of imaging data in three bands covering the northern galactic plane at latitudes $|b|<5^\circ$ and longitudes $30^{\rm o} \lesssim \ell \lesssim 220^{\rm o}$. Figure~\ref{fig:coverage} shows the distribution in the sky. The data have been obtained between August 2003 and December 2005 during a total of 212 nights (about 26 per cent of this observing time has been 
completely lost due to bad weather).

\begin{figure}
\centering
\includegraphics[trim=0 10 10 50,clip=true,width=\hsize]{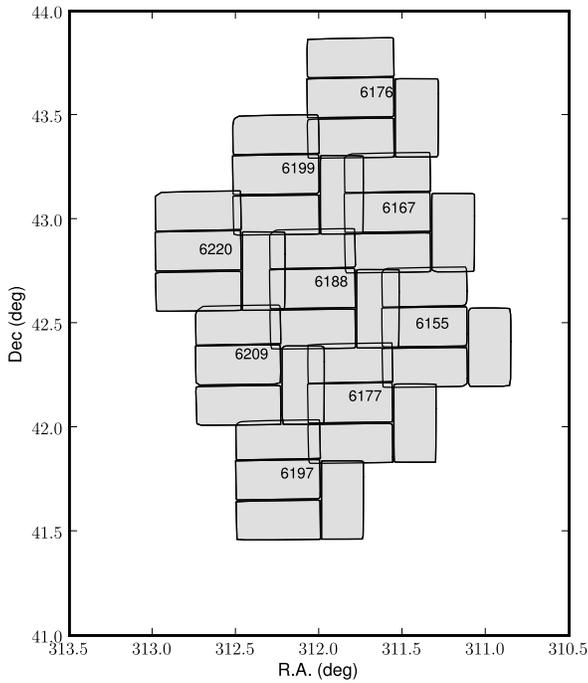}
\caption{Tiling configuration of $3\times3$ pointings covering an area of $\sim$2 square degrees. Chips are outlined in black. Field numbers are also shown. An additional pattern with an offset of $5'\times5'$ is not included in order to make the figure clearer.}
\label{fig:fields_obs}
\end{figure}

The area to be observed has been divided into ``fields'' each of them
corresponding to the coverage of the WFC.
Fields have been defined in a diagonal rather than rectangular
geometry in order to take advantage of the L-shape of the detector
configuration allowing for a 5 per cent overlap between adjacent
pointings. Figure~\ref{fig:fields_obs} provides an example of the
observation configuration.

\begin{figure}
\centering
\includegraphics[trim=10 0 10 10,clip=true,width=\hsize]{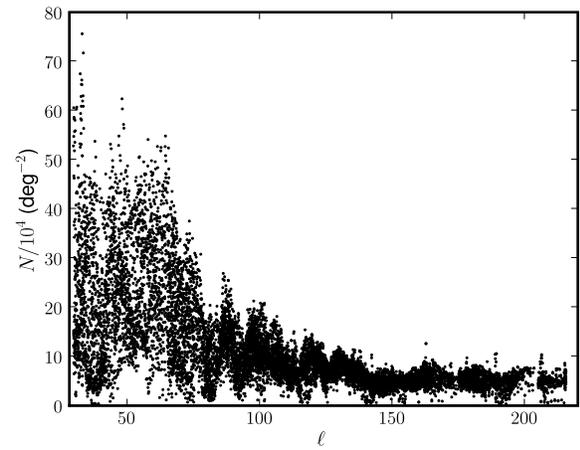}
\caption{Source counts as a function of Galactic longitude (one data point per IPHAS field).}
\label{fig:gcounts}
\end{figure}

Gaps between detectors are typically 20\arcsec. Therefore the area not
observed in each pointing due to the gaps is about 12 square arcmin
(i.e. $\sim$1 per cent). In order to cover the chip gaps and bad columns each
``field'' is observed in pairs which are referred to as ``on'' and
``off'' positions with an offset of $5\arcmin$W and $5\arcmin$S
between the pointings. Most of the objects are thus observed
twice. Field pairs are observed consecutively in the three bands
before moving to the next field pair in order to minimize the effects
of photometric variability. Fields with mean stellar seeing values
larger than 2 arcsec, ellipticity larger than 0.2, \rb-band sky counts
larger than 2000 ADUs (which indicates observations too close to the
moon or with clouds such that sensitivity is reduced to a
significantly lower magnitude limit) or anomalous star ratios between
the filters (which indicates variable clouds during observations) are
scheduled for re-observation in the three filters. Each field pair is
assigned an identification which we will refer to as ``Field
number''. The average stellar counts is $\sim$20,000 per square degree
although this value is highly dependent on the position along the
Galactic plane as seen in figure~\ref{fig:gcounts} which shows the number of
sources detected in each field as a function of Galactic
longitude.

\begin{figure}
\centering
\includegraphics[trim=10 0 30 30,clip=true,width=\hsize]{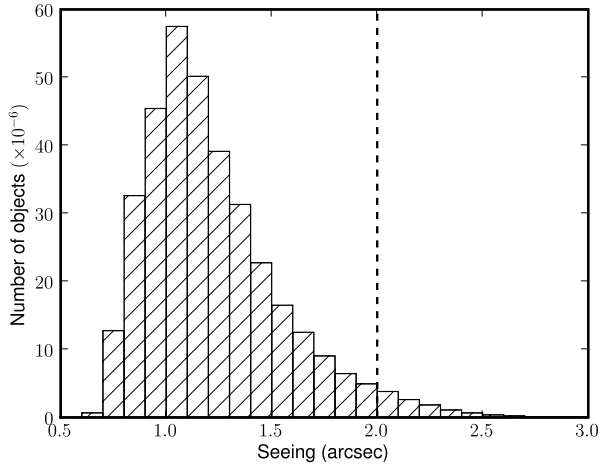}
\caption{Seeing distribution in the \rb band for the whole IDR. One of the criteria for inclusion in the \textbf{PhotoObjBest} catalogue is that objects have to be observed with seeing conditions better than 2.0 in the three bands (left of the dashed line).}
\label{fig:seeing}
\end{figure}

\begin{figure}
\centering
\includegraphics[trim=10 0 30 30,clip=true,width=\hsize]{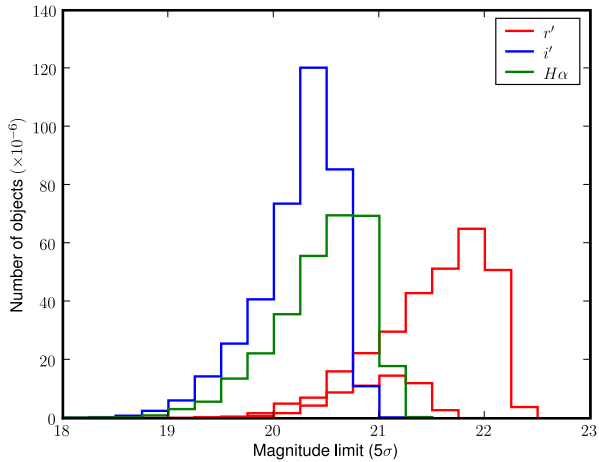}
\caption{Magnitude limit distribution for the  IDR in the three survey bands. The \rb\ band distribution has been split in two, the leftmost corresponding to the 2003 observations and the rightmost corresponding to post-2003 observations in which the exposure time was increased from 10\,s to 30\,s.}
\label{fig:maglim}
\end{figure}

The overall survey properties and quality of the data is summarised in table~\ref{tab:quality}.  The exposure time for each filter is 30\,s in \rb\ (except in the first year of observations, 2003, when it was 10\,s), 10\,s in \ib and 120\,s in \hb. All data obtained up to December 2005 have been included in the IDR regardless of seeing or photometric conditions. 
Figure~\ref{fig:seeing} shows the seeing values for these data in the \rb band. Although the mean value is about 1.26 arcsec (and the median 1.18 arcsec) there are clearly observations obtained in poorer conditions. The maximum acceptable seeing value for the data to be accepted as final survey quality is 2 arcsec in the three bands. Data with larger seeing values have been included in this release but they do not appear by default to the user. For this purpose we are releasing a full catalogue, \textbf{PhotoObj}, with all the observations obtained and a \textbf{PhotoObjBest} subset which only contains the observations which pass our scientific criteria for this release (see section~\ref{sec:catalogues} for more details). As noted in table~\ref{tab:quality} the coverage of this latter catalogue is about 1,000 square degrees and contains about 90 million unique objects. The Final Data Release of the data will extend the defined scientific criteria to the whole 1,800 square degrees. Figure~\ref{fig:maglim} shows the magnitude limit (5$\sigma$) for all
the objects in the three survey bands. Note that \rb\ observations obtained in 2003 are
shallower due to the smaller integration time used.  

\begin{table*}
\centering
\begin{tabular}{ll} \hline\hline 
\textbf{Parameter} & \textbf{Value} \\ \hline
Telescope & 2.5m Isaac Newton Telescope, La Palma \\
Instrument & Wide Field Camera \\
Field of view & 0.29 sq. deg. (per pointing) \\
Detector scale & 0.33 arcsec/pixel \\
Total area of survey & $\sim$1800 sq. deg \\
Total area in IDR & $\sim$1,600 sq. deg \\
Area included in the \textbf{PhotoObjBest} catalogue & $\sim$1000 sq. deg \\
Number of unique objects in the full catalogue & $\sim$200,000,000 \\
Number of unique objects in the  \textbf{PhotoObjBest} catalogue& $\sim$90,000,000 \\
Filters & \rb, \ib, \hb \\
Exposure time & 30\,s (\rb)$^\dagger$, 10\,s (\ib), 120s\, (\hb) \\
FWHM (median) & 1.3" (\rb), 1.2" (\ib), 1.3" (\hb) \\
Magnitude limit (Vega, 5$\sigma$) & 21.8 (\rb), 20.2 (\ib), 20.6 (\hb)\\
Astrometry & $<$0.1" (based on 2MASS) \\
\hline
\end{tabular}
\mbox{}\\
$^\dagger$ 10\,s in 2003
\caption{Survey parameters and IDR quality summary.}
\label{tab:quality}
\end{table*}

\section{Data Processing}
\label{sec:processing}

All the observations obtained with the INT telescope at La Palma are transmitted in pseudo-real time to the Cambridge Astronomy Survey Unit (CASU) for data archival via the internet. Although quasi real time data processing is possible it is preferable to perform the data processing on a run by run basis. The calibration is then performed using the whole run master flats and biases. For each observation the data are stored in a Multi-Extension Fits file (MEF) with a primary header describing the overall characteristics of the observation (pointing, filter, exposure time, ...) and four image extensions corresponding to each of the CCD detectors.

The data are processed by CASU as described in \cite{2001NewAR..45..105I} following the same steps as for the processing of the Wide Field Survey \citep{2005INGN....9....8I,2001NewAR..45...97M} which used the same telescope and camera combination. We provide here a short description of the reduction steps; additional details are available from \cite{2005MNRAS.362..753D}. 

The data are first debiassed (full 2D bias removal is necessary). Bad pixels and columns are then flagged and recorded in confidence maps, which are used during catalogue generation. Linearity tests using sequences of dome flats revealed that the CCDs have significant non-linearities at the level of 1 to 2 per cent depending on the detector and period of observation. Due to several changes to the WFC controllers the linearity values change from time to time so linearity measurements are updated periodically. A linearity correction using look-up-tables is then applied to all data. Flatfield images in each band are constructed by combining a series of sky flats obtained in bright sky conditions during the twilight for each run. 

The standard catalogue generation software \citep{1985MNRAS.214..575I}
makes direct use of the confidence maps previously generated (described in~\ref{cmaps})
 for a variety of tasks.  These include
    object detection and parametrization, producing quality control information, standard object 
descriptors and detected object overlay files.  The possibly varying sky 
background is estimated automatically, prior to object detection,
using a combination of robust iteratively clipped estimators.
The image catalogues are then further processed 
to yield morphological classification for detected objects and used to 
generate astrometric and photometric calibration information.

Each detected object has an attached set of descriptors, forming the
columns of the binary table and summarising derived position, shape and 
intensity information.  During further processing stages ancillary 
information such as the sky properties, seeing and mean stellar image 
ellipticity, are derived from the catalogues and stored in the 
FITS headers attached to each catalogue extension.  In addition to being
the primary astronomical products from the pipeline processing, the catalogues and associated derived 
summary information form the basis for astrometric and photometric 
calibration and quality control monitoring.

Finally an astrometric solution starts with a rough World Coordinate System (WCS) based on the
known telescope and camera geometry and is then progressively refined
using the 2MASS catalogue. The WFC field distortion is modelled
using a zenithal equidistant projection with polynomial radial distortion \citep[ZPN;
][]{2002A&A...395.1061G}. The resulting internal astrometric precision
is better than 100 mas over the whole WFC array (based on
intercomparison of overlap regions; see section~\ref{sec:astrom}).  
The object detection is performed in each
band separately using a standard APM-style object detection and
parametrization algorithm \citep{1985MNRAS.214..575I}.  The curve-of-growth,
defined as the flux inside an aperture as a function of its radius, is calculated 
measuring the flux in a set of apertures of radius $r/2$, $r$, $\sqrt{2}\,r$,
$2\,r$, $2\,\sqrt{2}\,r$ where $r=3.5$ pixels (1.2 arcsec) and an automatic
aperture correction (based on the average curve-of-growth for stellar
images) is applied to all detected objects. Finally a distortion correction is applied 
to the photometry to take into account the change of scale with respect to the optical 
axis distance (see subsection~\ref{sec:distcor}).

\begin{figure}
\includegraphics[width=\hsize]{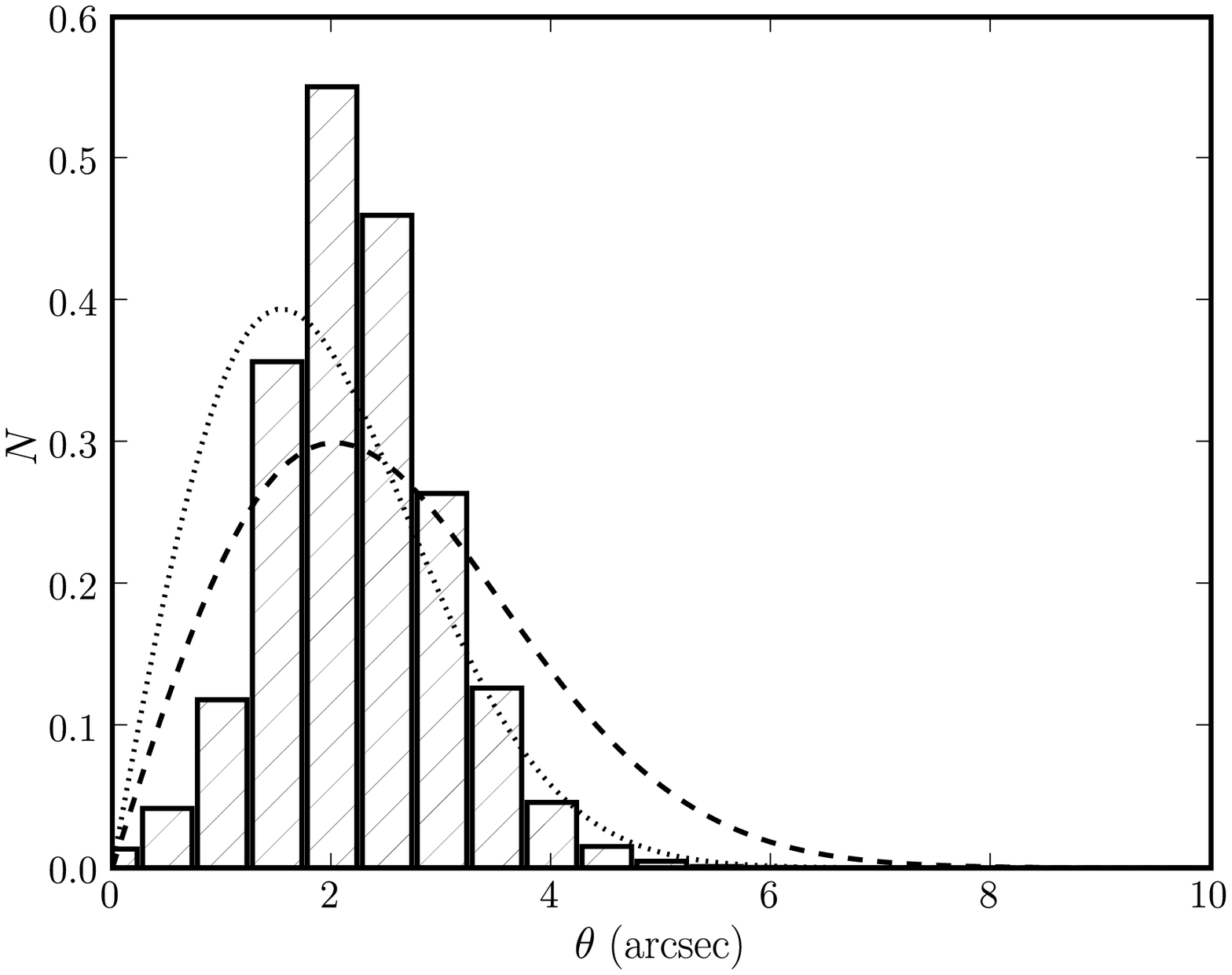} \\
\includegraphics[width=\hsize]{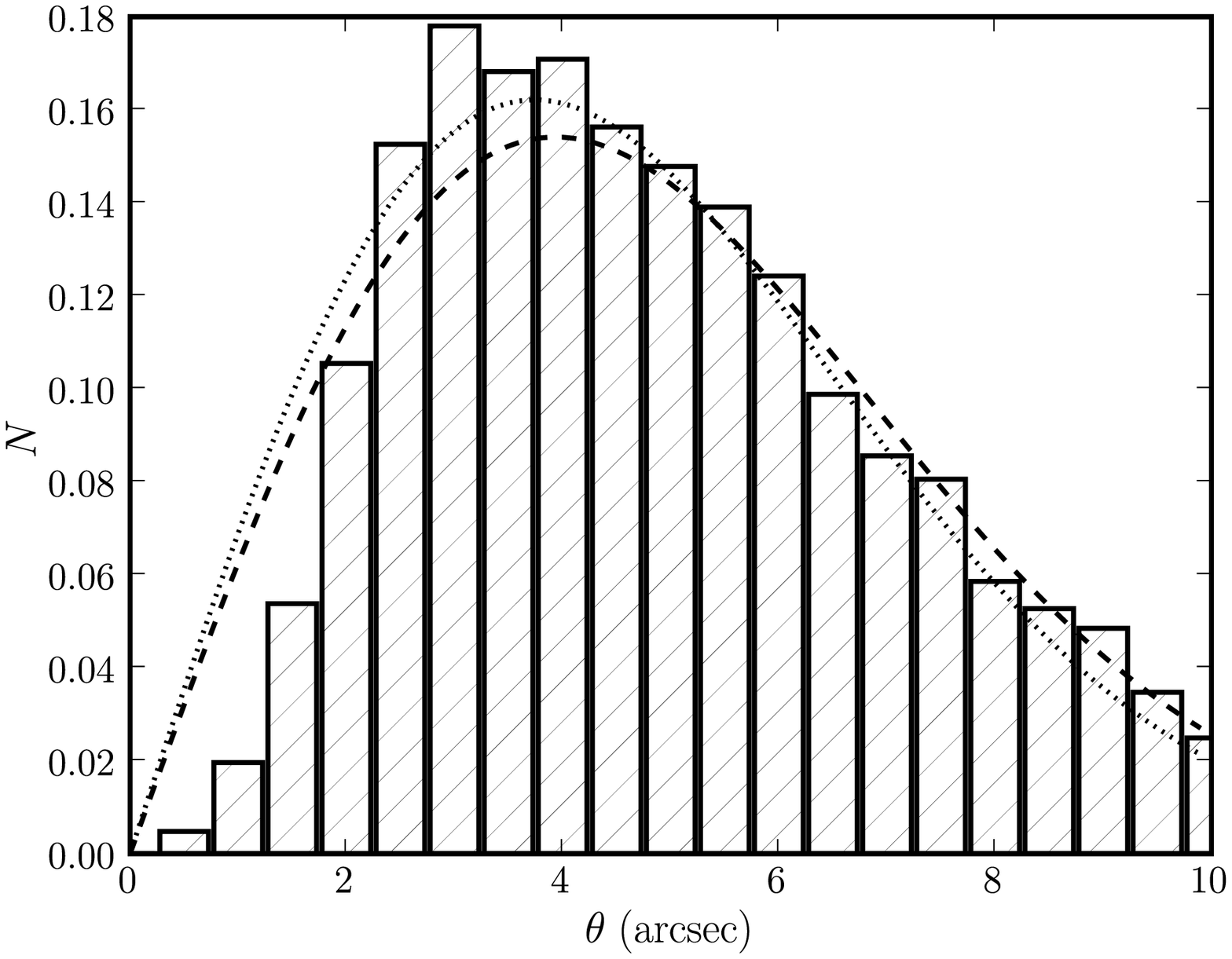}
\caption{Normalised nearest neighbour source distribution for a crowded field (top; 140 sources per arcmin$^2$, 146160 sources in total)
and for an average source density field (bottom; 40 sources per arcmin$^2$; 40760 sources in total) both observed with seeing of 1.2 arcsec. The dashed line shows the distribution for a random sample (equation~\ref{eq:nn}) using the observed source density while the dotted line shows the expectation when the average source density is corrected for incompleteness. Close pairs are missing due to the seeing and this is reflected in the smaller number of pairs with respect to the Poisson expectation at seeing smaller than 2 arcsec. This effect is larger in the crowded field leading to a large difference between the observed and expected normalized distributions.} 
\label{fig:neighbours}
\end{figure}

It is straightforward to show that an aperture of radius $\approx$FWHM  delivers 
optimal (i.e. $~$90 per cent of the total flux) photometry compared to more detailed PSF 
fitting \citep{1998MNRAS.296..339N} in non crowded areas. The default photometry is then done in an 
aperture of 1.2" and corrected for light loses outside the aperture. After 
adding some extra refinements to deal with the most crowded regions 
we find that there is little to be gained 
from PSF fitting (Irwin et al. 2008, in prep). The difference between crowded and non crowded areas is shown in figure~\ref{fig:neighbours} which represents
the nearest neighbour source distribution for two different fields: one with high 
source density to illustrate the worst case scenario of a crowded field and another
with an average source density. For comparison we use the nearest neighbour probability
density distribution (PDF) expected for a random sample which can be written
as \citep{1981ApJ...246..122B}
\begin{equation}
P_{NN}(\theta) = 2\,\pi\,\rho\,\theta\,e^{-\pi\rho\theta^2} \label{eq:nn}
\end{equation}
where $\rho$ is the average density and $\pi\rho\theta^2$ is the number of
neighbours in the interval $(0, \pi\theta^2)$. The expected mean nearest neighbour separation
can then be calculated as \citep{1989MNRAS.241..109S}
\begin{equation}
\langle\,\theta\,\rangle = \int_0^\infty \theta\,P_{NN}(\theta)\,d\theta = \frac{1}{2\sqrt{\rho}}
\end{equation}
For the space densities of the fields plotted in figure~\ref{fig:neighbours} we obtain 
$\langle\,\theta\,\rangle=2.5"$ for the crowded field and $\langle\,\theta\,\rangle=4.7"$ for the average density field. The maximum of the $P_{NN}$ distribution is given by $\theta_{m} = 1 / \sqrt{2\pi\rho}$ and corresponds to $\theta_m=2.0"$ and $\theta_m=3.8"$ in the crowded and average density fields respectively. The large discrepancy between the random and observed distributions for the
high density field at $\theta>2"$ is due to missing close pairs which produces an overestimation
of the nearest neighbours at $\theta\simeq2"$. We can calculate the fraction of missed objects
due to crowding using the following expression \citep{1984AJ.....89...83I}
\begin{equation}
\rho_{\rm corr} = - \frac{log (1 - 4 \pi \rho \, \theta_{FWHM}^2)}{4 \pi \, \theta_{FWHM}^2} \label{eq:avcorr}
\end{equation}
where $\rho_{\rm corr}$ is the corrected average density and $\theta_{FWHM}$ the FWHM of the images(i.e. 1.2 arcsec). For the average density field the fraction of sources lost due to close pairs is 9 per cent while in the crowded field this percentage increases to 41 per cent. Note as well that non resolved objects pair, on average, with objects at a distance $\sim \langle\,\theta\,\rangle$ occupying, and overestimating, the central bins of the nearest neighbour distribution (figure~\ref{fig:neighbours} top). We also plot in figure~\ref{fig:neighbours} the corrected nearest neighbour PDF as a dotted line using the corrected average density given by equation ~\ref{eq:avcorr}.

Note that the crowded field case represents a extreme case which has source density more than three times higher than the median source density of the survey.

The values obtained for the mean nearest neighbour separations are clearly well 
above our seeing values in both cases indicating that aperture photometry is appropriate for our case.

\subsection{Morphological Classification}

The morphological classification is based primarily on the aperture fluxes
and the discrete curve-of-growth for each detected object defined by them.  
Ancilliary information such as object ellipticity derived from 
intensity-weighted second moments is also used but only as a secondary 
indicator.  The curve-of-growth is a direct measure of the integral of 
the point spread function (PSF) out to various radii and is independent of 
magnitude if the data are properly linearised, and if saturated 
images are excluded.  In using this property the classifier further assumes 
that the effective PSF for stellar objects is constant over each 
detector,\footnote{In practice we find that the effects of the spatial 
variation of the PSF on the aperture fluxes at the detector level are 
generally negligible.} although individual detectors are allowed to have 
different PSFs since all detectors, pointings and passbands are treated 
independently.

\begin{figure}
\includegraphics[width=0.75\hsize,angle=-90]{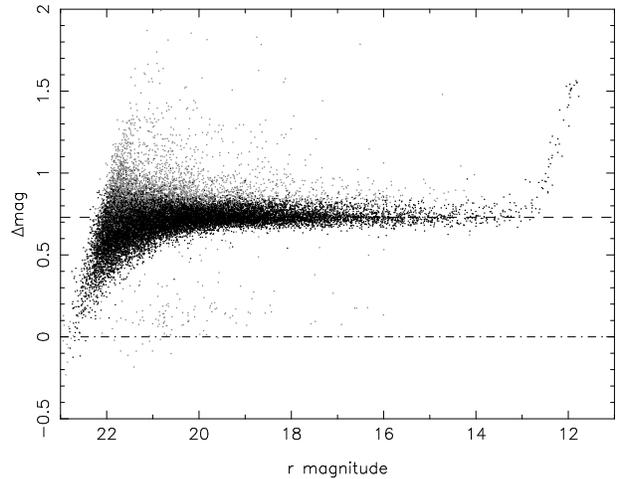}
\caption{An example of how the difference in flux recorded between
pairs of apertures is used to provide information for morphological
classification.  The stellar locus (black points) is self-defining at a
constant offset (here 0.73 magnitudes) while the spread of points about
this locus, after iterative clipping, defines the average scatter of
stellar objects as a function of magnitude.  Grey points denote noise-like
objects that lie well below the stellar locus ridge, or non-stellar or
blended stellar objects which lie above the stellar population.  Saturated
stars form the upturned sequence brighter than 13th magnitude. The offset
of the top dashed line defines the average differential aperture correction
for stellar objects for this pair of apertures and in combination with other
pairs of apertures defines the overall stellar curve-of-growth and hence
individual stellar aperture corrections.}
\label{fig:classification}
\end{figure}

The average stellar locus on each detector in these parameter spaces is 
generally well-defined and is used as the basis for a null hypothesis 
stellarness test.  Figure~\ref{fig:classification} shows an example of this comparing the 
magnitude difference recorded in two apertures (1.2 and 2.4 arcsec diameter) 
as a function of the derived overall magnitude and illustrates how this
relates to the final classification for this frame.

In practice, the aperture with radius $r=1.2$ arcsec is used as a fixed 
reference and also defines the internal magnitude (flux) scale.  The linearity 
of the system implies that the position of the stellar locus for 
{\it any function} of the aperture fluxes is independent of magnitude (at 
least until images saturate). Therefore marginalising the flux ratios over 
magnitude yields one-dimensional distributions that can be used to greatly 
simplify locating the stellar locus using an iteratively clipped median.  
With the location estimated, the median of the absolute deviation from the 
median (MAD) provides a robust measure of the scatter about this locus as a 
function of magnitude, at least until galaxies dominate in number.  This 
process is repeated iteratively for each distribution, using 3-sigma clipping 
to remove non-stellar outliers, until satisfactory convergence is reached.
After convergence the equivalent Gaussian sigma is estimated using 
$\sigma_{gauss} = 1.48 MAD$ and by this means each of the image shape 
descriptors (in this case flux ratios or ellipticity) can be renormalised to 
follow a zero-median, unit variance Gaussian-like $N(0,1)$ distribution. 

These measures are then directly combined to form the final classification
statistic. The combination (essentially a unweighted sum of the normalised 
signed distributions) is designed to preserve information on the ``sharpness'' 
of the object profile and is finally renormalised, as a function of magnitude, 
to produce the equivalent of an overall $N(0,1)$ classification statistic.

In practice measures derived from real images do not exactly follow Gaussian
distributions.  However, by combining multiple normalised distributions 
(with well-defined 1st and 2nd moments), the Central Limit Theorem works 
in our favour such that the resulting overall statistic is Gaussian-like to
a reasonable approximation and hence can be used with due care as the 
likelihood component of a Bayesian Classification scheme, making optional 
use of prior knowledge.

Objects lying within 2--3$\sigma$ of the stellar locus (\ie of zero) are 
generally flagged as stellar images, those below --3 to --5$\sigma$ 
(\ie sharper) as noise-like, and those above 
2--3$\sigma$ (\ie more diffuse) as non-stellar.
Although the discrete classification scheme is based on the $N(0,1)$ measure
of stellar appearance it also has several overrides built in to attempt to make 
it more reliable.  For example, adjustments to the boundaries at the faint-end 
(to cope with increased $rms$ noise in the statistic) and at the bright-end (to 
cope with saturation effects) are also made, while the overall image
ellipticity provides a further check.

A by-product of the curve-of-growth analysis and the classification is an
estimate of the average PSF aperture correction for each detector for those 
apertures (up to and including $4r$, which includes typically $\sim$99 per cent, 
or more, of the total stellar flux) used in deriving the classification 
statistic.  Accurate assessment of the aperture correction to place the 
(stellar) fluxes on a total flux scale is a crucial component of the overall 
calibration.  We find that this method of deriving aperture corrections 
contributes $\leq \pm$1 per cent to the overall photometry error budget and also 
provides a useful first order seeing correction for non-stellar sources.  
Further by-products of the morphological classification process are improved 
estimates of the seeing and average PSF ellipticity from making better use 
of well-defined stellar-only sources.  These parameters are required for 
quality control monitoring of telescope performance and ``atmospheric'' 
seeing.

\subsection{Photometric Calibration}

Photometric calibration is done using series of Landolt standard
stars \citep{1992AJ....104..340L} with photometry in the SDSS system with additional stars from \cite{2000PASP..112..925S}. For each
night a zero point and error estimate using the observations of all the standard fields in each filter is derived. For photometric nights
the calibration over the whole mosaic has an accuracy of 1-2 per cent.
For the purpose of the photometric calibration, standard star observations have been obtained each night at an interval of 2\,h and have been used to calibrate the \rb and \ib frames. The \hb frames have been calibrated using a fixed zero-point offset of 3.14 magnitudes with respect to \rb, corresponding to the magnitude difference in $r'-$\hb for Vega. This value has been calculated by convolving the HST Vega spectrum \citep{2007ASPC..364..315B} with the atmospheric throughput, INT optical throughput (primary mirror and field corrector), CCD efficiency and the WFC filter curves (available from the ING web pages at www.ing.iac.es)
so that $r'-H\alpha=0$ for Vega.

All calibration is by default corrected during pipeline processing for the mean atmospheric
extinction at La Palma (0.09 in \rb and \hb and 0.05 in \ib per unit airmass).

\begin{figure}
\includegraphics[width=0.7\hsize,angle=-90]{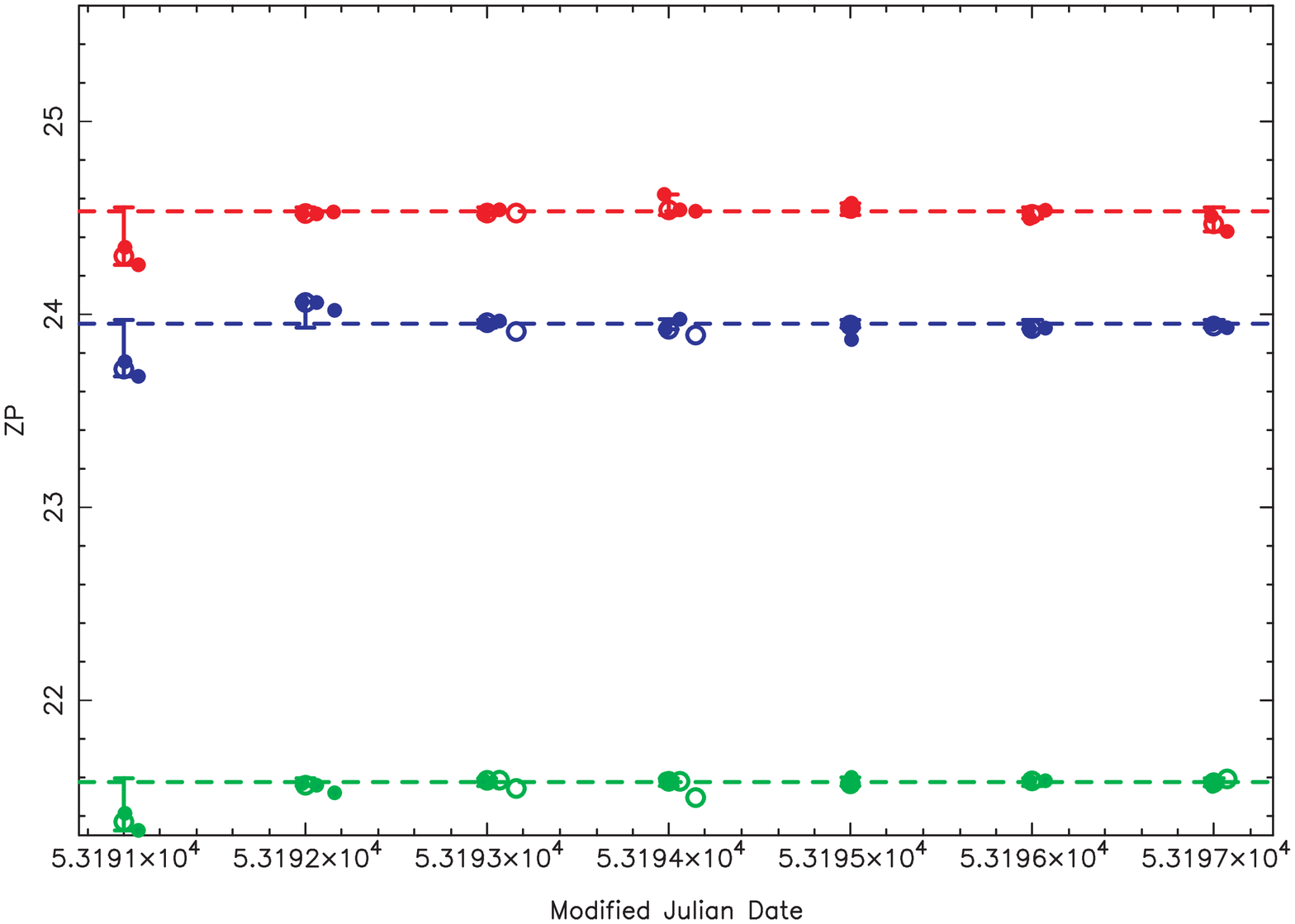} \\ 
\includegraphics[width=0.7\hsize,angle=-90]{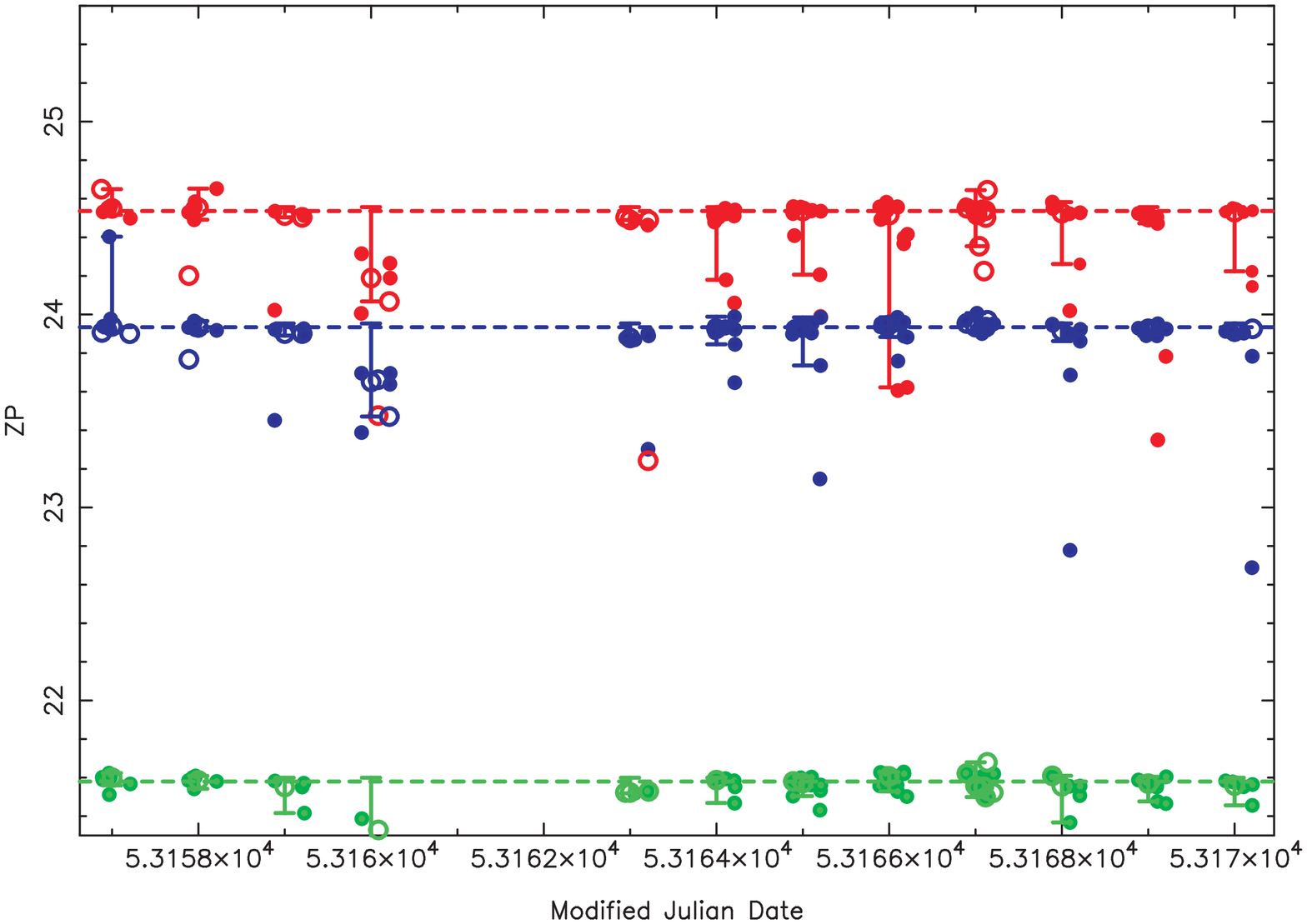} 
\caption{Photometric zero point (ZP) versus date for two runs. The dashed lines are the expected values for a good photometric night in \rb\ (red), \ib\ (blue) and \hb\ (green). Dots represent the individual measurements for each frame. The top figure shows a good photometric run in which all the zero points are consistent with the expected values. The bottom one instead shows a variable run in which a large percentage of observations have been obtained in bad photometric conditions and therefore not included in the \textbf{PhotoObjBest} catalogue. These plots are used to flag problematic frames and runs.}
\label{fig:zp}
\end{figure}

\begin{figure}
\centering
\includegraphics[trim=40 30 40 40,clip=true,width=\hsize]{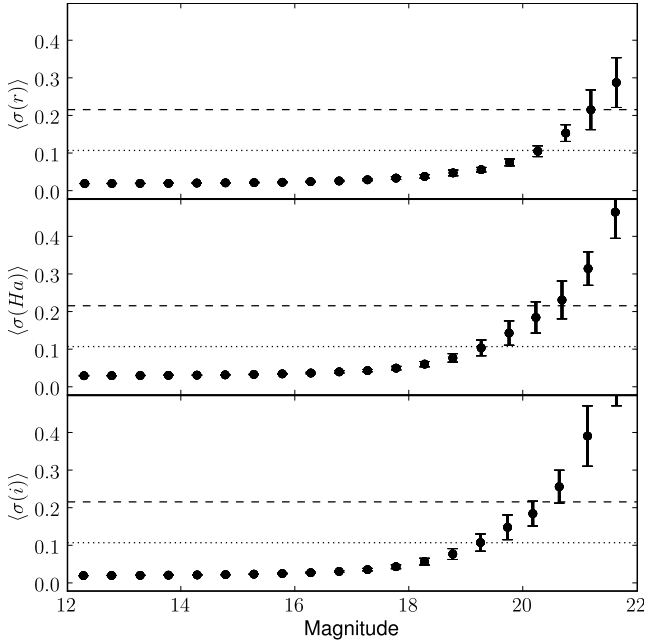}
\caption{Photometric measurement uncertainties as a function of magnitude in 0.5 mag bins for all the objects
included in \textbf{PhotoObjBest}. The vertical error bars show the rms dispersion in the mean uncertainties in each brightness bin and the horizontal lines in each panel show the SNR=10 (dotted) and SNR=5 (dashed) values.}
\label{fig:photomsigma}
\end{figure}

\begin{figure}
\includegraphics[width=\hsize]{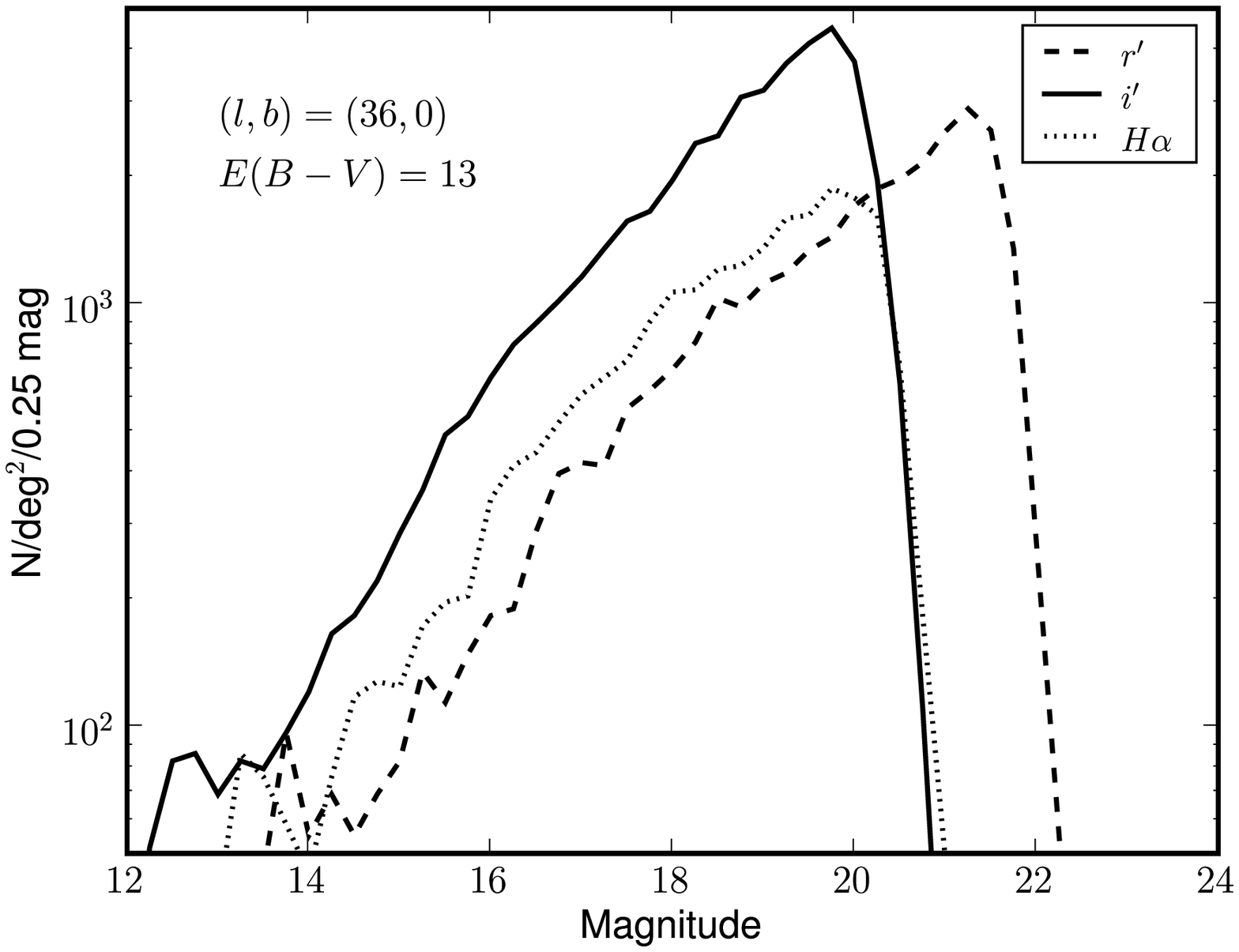} \\
\includegraphics[width=\hsize]{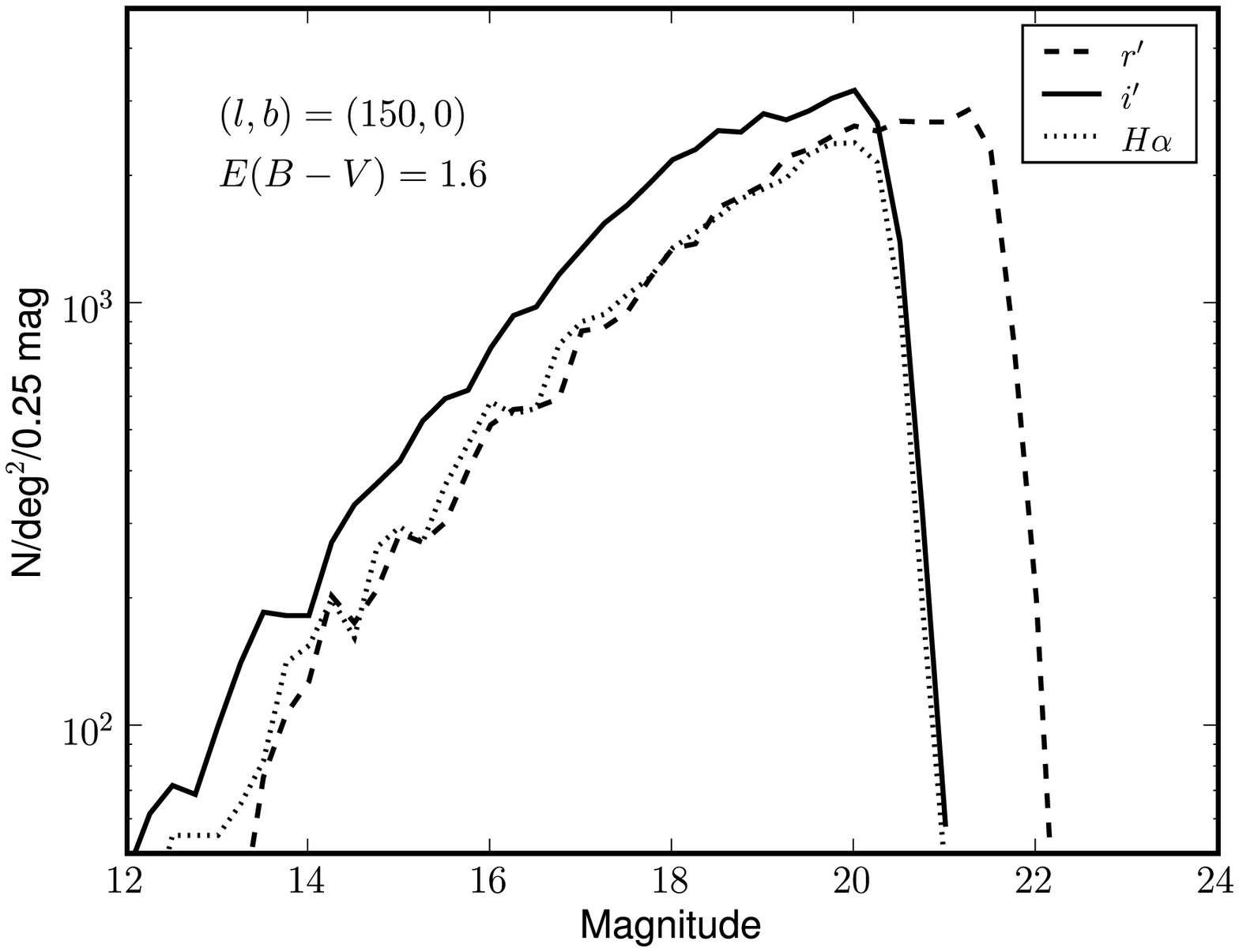}
\caption{Star counts along two contrasted lines of sight with star densities of 
$\sim$50,000 deg$^2$ and different reddening.}
\label{fig:starcounts}
\end{figure}

During non-photometric nights, in otherwise acceptable observing conditions,
we find that the derived zero point systematic errors can be up to 10 per cent
or more.  Although the pipeline usually successfully flags such nights
as non-photometric (see figure~\ref{fig:zp}) it still leaves open the problem of what to do
about tracking the varying extinction during these nights. For this Initial Data Release we have not included a global photometric solution for the whole survey. 
While it is possible to perform an offset to a common reference system for a small subset of the observations
using overlaps, as demonstrated below in section~\ref{sec:science}, the application of such methods to the
whole survey area is more complicated since one has to carefully select good reference frames and
allow for variation in parameters like extinction or zero point over the whole survey period. Availability of good photometric observations to tie the solution as much as possible is crucial. Clearly this
is a major task which we are currently investigating and is best accomplished when the survey is complete.
Data from non-photometric nights  have been included in the release but flagged as such so that they do not appear in our \textbf{PhotoObjBest} catalogue. Figure~\ref{fig:photomsigma} shows the mean magnitude rms for the objects included in this catalogue, zero point uncertainties are
at the level of 2 per cent for \rb\ and \ib\ and 3 per cent for \hb. We also plot in figure~\ref{fig:starcounts} the stellar counts in the three bands along two different line of sights in the Galactic plane with different reddening.

All magnitudes quoted are in the Vega system but we give the conversion factors to the AB system in table~\ref{tab:filters}.

\subsection{Astrometric Calibration} 
\label{sec:astrom}

Astrometric calibration is a multi-stage process and aims to provide each 
image, and any derived catalogues, with a World Coordinate System (WCS) to 
convert between pixel and celestial coordinates. This happens in the pipeline 
in two generic stages.  

An initial WCS based on knowledge of the instrument, \eg orientation, 
field-scale, telescope pointing, is embedded in the FITS headers, with 
telescope-specific information in the primary header and detector-specific 
information in the secondary headers.  This serves to locate each detector 
image to within a few to several arcsec, depending on the pointing accuracy 
of the telescope and model parameters.  The essential information required
is the RA and Dec of the pointing, a (stable) reference point on the detector 
grid for those coordinates (\eg the optical axis of the instrument), 
the central pixel scale, the rotation of the camera, the relative 
orientation of each detector and the geometrical distortion of the 
telescope and camera optics, which defines the astrometric projection to use.

\begin{figure}
\centering
\includegraphics[trim=20 0 0 0,clip=true,width=\hsize,angle=-90]{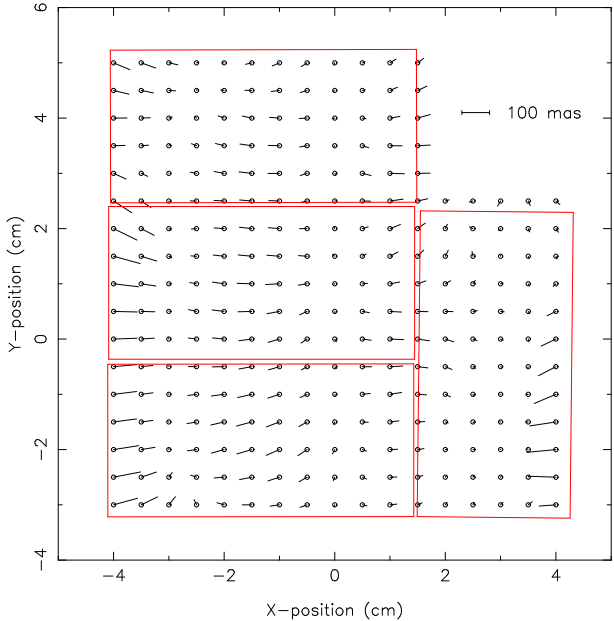}
\caption{Astrometric residuals derived from a series of independent WFC pointings. The residuals are shown in physical coordinates with respect to the center of the optical axis.}
\label{fig:astrom}
\end{figure}

Given a rough WCS for the processed frames, a more accurate WCS can be 
defined using astrometric standards.  We have based our calibration
on the 2MASS point source catalog \citep{2006AJ....131.1163S} for several reasons: 
it is an all-sky NIR survey; it is calibrated on the International Celestial 
Reference System (ICRS); it provides at least 100 or more suitable standards 
per pointing; it is a relatively recent epoch (mid-1990s) minimising proper 
motion problems; the global systematics are better than 100 mas over the 
entire sky \citep{2003IAUJD..16E..43Z}; and for 2MASS point sources with 
signal:to:noise $>$10:1 the $rms$ accuracy per source is $\simlt$100 mas.

By stacking the astrometric residuals from a series of independent pointings and CCD WCS solutions it is possible to assess the accuracy of the simple INT distortion model. This is illustrated in the figure~\ref{fig:astrom} using the average residuals from a stack of a one week run. The end product of the full pipeline currently has an astrometric precision better than 100 mas over the whole array (ie. across CCDs), as determined by analysis of independently calibrated adjacent overlapping pointings.

\section{Data Products}
\label{sec:dataproducts}

\subsection{Images}
\label{sec:distcor}

There are two types of imaging data products: (1) object images and (2) confidence maps. At this point we do not provide stacked images from the overlapping frames. We describe each of them separately.

As stated before, each image is a MEF file containing the four CCD frames. Each CCD is photometrically and astrometrically calibrated as described in the previous section. Photometry can be extracted from the images using the following expression in the instrumental system
\begin{equation}
  m = {\rm ZP} - \kappa \, (X - 1) - 2.5 \, \log_{10} ( f / t )  + c  \label{eq:mags}
\end{equation}
where $m$ is the calibrated magnitude, ZP is the zero point at unit airmass as extracted from the image headers, $\kappa$ is the extinction coefficient, $X$ is the airmass, $f$ is the flux in counts and $t$ is the exposure time. 
The $c$ correction is based on the median dark sky recorded in science frames compared to the median for all the CCDs and as such is an ancillary correction to the gain correction derived from the flatfield (usually twilight flats). This correction is in the header keyword \verb+PERCORR+ and is typically smaller than 1 per cent and also corrects for the small quantum efficiency differences between the detectors.

Astrometric information is written in the FITS headers using a ZPN projection which allows for a description of the non uniform scale over the field of view without performing pixel resampling (necessary e.g. in the case of describing the distortions using a more usual tangential projection). 

\begin{figure}
\centering
\includegraphics[width=\hsize]{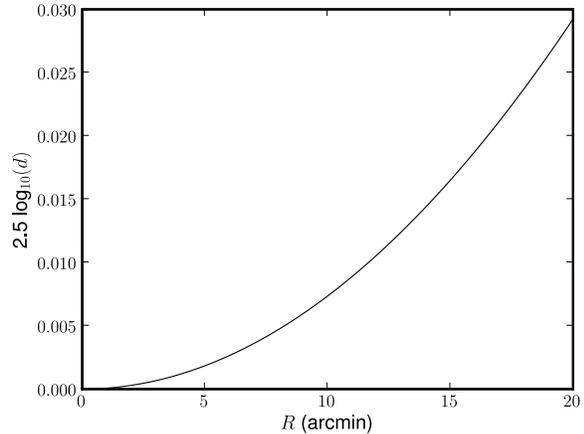} 
\caption{Effect of scale change in photometry (magnitude versus distance to field centre).}
\label{fig:distortion}
\end{figure}

The aim of conventional flatfielding is to create a flat background by normalising out perceived variations from uniformly illuminated frames. If the sky area per pixel changes then this is reflected in a systematic error in the derived photometry. However the change of scale, i.e., the change of sky area per pixel, also creates photometric effects. The magnitude of the multiplicative correction to be applied to the measured flux can be
modelled as (Irwin et al. 2008, in prep)
\begin{equation}
d = \left( 1 + \frac{3 \, P_3 \, R^2}{P_1} \right) \left(1 + \frac{P_3 \, R^2}{P_1} \right)
\label{eq:distcor}
\end{equation}
where $R$ is the distance from the optical axis and for the WFC the coefficients are
$P_1=1$ and $P_3=220$ (corresponding to the \verb+PV2_1+ and \verb+PV2_3+ WCS keywords as defined in \cite{2002A&A...395.1061G}). Figure~\ref{fig:distortion} shows the effect of this term in the photometry which is at a maximum level of only 2 per cent at the outer parts of the frames.
The value displayed in the figure has to be subtracted from the magnitude calculated in equation~\ref{eq:mags} 
and is automatically included in the catalogue data products but has to be applied by the user when measuring fluxes from the images. Table~\ref{tab:keywords} lists the main
photometric keywords present in the image headers and needed to convert the fluxes
to magnitudes as explained in this section.

\begin{table}
\centering
\begin{tabular}{ll} \hline\hline
\textbf{Keyword} & \textbf{Description} \\ \hline
MAGZPT & [mag] Magnitude zero point for default extinction \\
                & measured at unit airmass \\
MAGZRR & [mag] Error in magnitude zero point \\
AIRMASS & Airmass at start of observation \\
EXPTIME & [sec] Exposure time \\
WFFBAND & Filter bandpass \\
SKYLEVEL & [counts/pix] Median sky brightness \\
SKYNOISE & [counts] Pixel noise at sky level\\
SEEING & [pixels] Mean stellar FWHM \\
APCOR & [mag] Stellar aperture correction \\
PERCORR & [mag] Sky calibration correction \\
NUMBRMS & Number of objects used when computing \\
                   & the astrometric fit \\
STDCRMS & [arcsec] Astrometric fit error \\
\hline
\end{tabular}
\caption{Subset of relevant photometric and astrometric keywords available from the image headers after pipeline processing.} \label{tab:keywords}
\end{table}

\subsubsection{Confidence maps} \label{cmaps}

We define a confidence $c_{ij}$ map, where $j$ refers to pixel $j$ of frame $i$, that is an inverse variance weight map in units of
relative variance denoting the "confidence" associated with the
flux value in each pixel.  The $c_{ij}$ are normalised to a median
level of 100 per cent (i.e. $\langle c_{ij} \rangle = 1$).
This has the advantage that the same map can also be used to 
encode for hot, bad or dead pixels, by assigning zero confidence. 

The confidence map for each frame is derived from analysis of the flat fields and is unique for each filter/detector combination. It encodes individual pixel sensitivities and allows for vignetted or bad regions to be masked. These confidence maps are used in the object detection phase of the processing.

Use of the confidence maps for weighted co-addition of frames, or for object
detection, simply requires an overall estimate of the average noise 
properties of the frame. This can be readily derived from the measured sky 
noise, in the Poisson noise-limited case, or from a combination of this and 
the known system characteristics (\eg gain, readout noise). All processed frames have 
an associated derived confidence map which is propagated through the 
processing chain. A detailed description of the process is given in 
Irwin et al. 2008, in prep.

\subsection{Catalogues}
\label{sec:catalogues}

The object catalogue is made available  using a relational database system. We have followed the philosophy of reducing  the 
number of tables presented to the user as much as possible. The database contains four tables which are summarised below.

\begin{itemize}
  \item \textbf{FieldsAll} (table~\ref{tab:fieldsall}). Information about the field centres and number of objects detected. Each field pair has associated a unique field number (\verb+fieldno+) and each of those an on/off position (\verb+onoff+ where a value of 0 has been assigned to the off position and 1 to the on position). These two are combined into the \verb+fieldID+ to provide a unique identifier for each pointing. Equatorial and galactic coordinates for each pointing as well as number of detections are also provided.
  \item \textbf{ChipsAll} (table~\ref{tab:chipsall}). This table contains parameters about the observation (airmass, exposure time, World Coordinate System) as well as the Data Quality Control attributes for each chip, i.e, mean seeing and ellipticity, zero point, sky level, etc. Also for each pointing the Galactic coordinates are available as well as the Galactic extinction as measured from the Schlegel maps~\citep{1998ApJ...500..525S} using a 4 pixel linear interpolation for each position.
  \item \textbf{PhotoObj} (table~\ref{tab:photoobj}). Main catalogue containing all detections with image derived parameters and calibrated magnitudes. Each row has a set of keys (\verb+chip_r_id+, \verb+chip_i_id+, \verb+chip_ha_id+, \verb+field_id+) which allow the user to enquire about the photometric conditions (via the \textbf{ChipsAll} table) under which the measurements of a particular object have been done or to which field it belongs. Additional flags (e.g \verb+seeing_flag+) allow the selection of objects observed in particular (seeing) conditions. We have also included the Hierarchical Triangular Mesh index (\verb+htmid+) as defined by the Sloan Digital Survey Science Archive~\footnote{see http://skyserver.org/htm/}. The total number of detections in this catalogue is 411 million including duplicate observations. Due to the on/off observation strategy which leads to every source being observed twice this means about 200 million unique objects.
  \item \textbf{PhotoObjBest} (table~\ref{tab:photoobj}). Subset of \textbf{PhotoObj} with additional constraints used to select only those observations obtained in good photometric conditions and thus will vary little between the IDR and the Final Data Release. The data obtained in the Aug 2003 and Dec 2004 observing runs have been discarded due to poor photometric conditions. This catalogue contains only observations for those nights for which the mean magnitude zero point rms in the \rb\ band is better than 5 per cent, the mean \rb\ zero point is greater than 24.2 and the individual frames have been obtained in seeing conditions better than 2 arcsec and ellipticity better than 0.2. Additionally only sources detected in all three bands have been included. The total number of detections in this catalogue is 181 million (approximately 90 million unique objects).
  \item \textbf{tGetNearbyObj}. This is a store procedure to perform a cone search without having to use the complete haversine formula explicitly to calculate angular distances. Given the coordinates (\verb+ra+, \verb+dec+) in degrees and a radius (\verb+r+) in arcmin it returns a list of identifiers for the objects in such defined cone. The output can then be joined with one of the \textbf{PhotoObj/Best} tables to return the object parameters (see example in appendix A).
\end{itemize}

Tables ~\ref{tab:fieldsall}, \ref{tab:chipsall} and \ref{tab:photoobj} list the column names and their description. Columns marked with an asterisk are indexed by the database software, i.e., searches in those columns are particularly fast. Users should use when possible at least one indexed column in order to speed up the queries.

\begin{figure}
\includegraphics[width=\hsize]{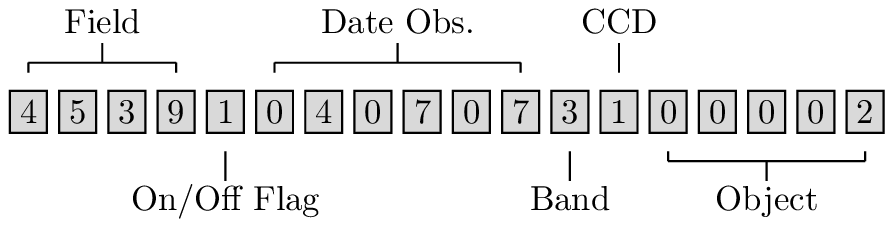}
\caption{The unique object ID encodes the field number, on/off flag (1=on, 0=off), date of observation (in the format YYMMDD), band (1=\rb; 2=\ib; 3=\hb) and CCD number.}
\label{fig:objid}
\end{figure}

Each entry in each table is assigned a unique ID number used for cross reference. Figure~\ref{fig:objid} shows an object ID in the \textbf{PhotoObj/Best} tables and how it is decoded. Each entry in the \textbf{ChipsAll} table is encoded in the same way but without the last five figures which represent the object number.

Individual detection star/galaxy classifications are combined into one single value following a similar approach to the WFCAM Science Archive \citep[WSA;][]{2007arXiv0711.3593H}. A classification table is defined which assigns probability values for each classification code and then each classification is combined for a merged source using Bayesian classification rules assuming each value is independent:

\begin{equation}
P(c_k) = \prod_j P(c_k)_j / \sum_k \prod_j P(c_k)_j
\end{equation}

where $c_k$ is the classification flag and $i$ denotes the $i^{\rm th}$ single detection passband measurement available. Decision thresholds for the resulting discrete classification flag are 90 per cent for definitive and 70 per cent for probable. An additional decision rule enforces class\_flag=255 (saturated) when any individual classification flag indicates saturation.

\begin{table}
\centering
\begin{tabular}{lll}
\hline
\textbf{Column Name} & \textbf{Type} & \textbf{Description} \\ \hline
fieldID* & long & Field ID \\
onoff* & int & on/off observation flag \\
fieldno* & int & Field number \\
ra* & float & R.A. of centre of field (deg, J2000) \\
dec* & float & Dec of centre of field (deg, J2000) \\
glon & float & Galactic longitude of centre of field (deg)\\
glat & float & Galactic latitude of centre of field (deg)\\
nobj & int & Number of objects detected  \\ \hline
\end{tabular}
\caption {Column names, types and description of FieldsAll table. Columns marked with an asterisk are indexed.}
\label{tab:fieldsall}
\end{table}

\begin{table}
\centering
\begin{tabular}{lll} \hline
\textbf{Column Name} & \textbf{Type} & \textbf{Description} \\ \hline
chipID* & long & chip ID \\ 
runno* & int & Run number \\
ra* & float & R.A. of the chip center (deg, J2000) \\
dec* & float & Dec of the chip center (deg, J2000) \\
glon & float & Galactic longitude (deg) \\
glat & float & Galactic latitude (deg) \\
ebv & real & $E(B-V)$ from Schlegel dust map \\
naxis1 & int & Chip size along x axis (pixels)\\
naxis2 & int & Chip size along y axis (pixels)\\
mjd & float & MJD of observation \\
magzpt* & real & Photometric zero point (Vega) \\
magzrr & real & Zero point error (Vega)\\
extinct & real & Atmospheric Extinction \\
airmass* & real & Airmass \\
exptime & real & Exposure time (seconds)\\
night* & int & Night of observation (YYYYMMDD)\\
dateobs* & int & Date of observation (YYYYMMDD)\\
band* & char & Filter name \\
maglim* & real & Magnitude limit (5$\sigma$, Vega)\\
seeing* & real & Mean seeing (arcsec)\\
elliptic* & real & Mean stellar ellipticity \\
cd1\_1 & float & CD Matrix \\
cd1\_1 & float & CD Matrix \\
cd1\_1 & float & CD Matrix \\
cd1\_1 & float & CD Matrix \\
crval1 & float & R.A. of reference pixel (deg) \\
crval2 & float & Dec of reference pixel (deg) \\
crpix1 & float & x coordinate of reference pixel\\
crpix2 & float & y coordinate of reference pixel\\
percorr & real & Sky correction (Vega)\\
apcor & real & Aperture correction (Vega)\\
skylevel & real & Sky level (counts)\\
skynoise & real & Sky noise (counts)\\
nobj & real & number of detected objects\\
field\_id* & long & link to FieldsAll table \\ \hline
\end{tabular}
\caption {Column names, types and description of data quality control table ChipsAll. Columns marked with an asterisk are indexed.}
\label{tab:chipsall}
\end{table}

\begin{table}
\centering
\begin{tabular}{lll}
\hline
\textbf{Column Name} & \textbf{Type} & \textbf{Description} \\ \hline
objID* & long & Object ID \\
objname & char & Object Name \\
ra* & float & R.A. in degrees (deg, J2000) \\
dec* & float & Dec in degrees (deg, J2000) \\
glon & float & Galactic longitude (deg) \\
glat & float & Galactic latitude (deg) \\
coremag\_r & float & \rb aperture corrected magnitude \\
coremag\_i & float & \ib aperture corrected magnitude \\
coremag\_ha & float & \hb aperture corrected magnitude \\
coremagerr\_r & float & \rb magnitude error \\
coremagerr\_i & float & \ib magnitude error \\
coremagerr\_ha & float & \hb magnitude error \\
rmi* & float & \rb - \ib colour \\
rmha* & float & \rb - \hb colour \\
rmierr & float & \rb - \ib colour error \\
rmhaerr & float & \rb - \hb colour error \\
class\_r & int & \rb star/galaxy clasification index \\
class\_i & int & \ib star/galaxy clasification index \\
class\_ha & int & \hb star/galaxy clasification index \\
ellipt\_r & real & ellipticity as measured in \rb \\
ellipt\_i & real & ellipticity as measured in \ib \\
ellipt\_ha & real & ellipticity as measured in \hb \\
posang\_r & real & position angle as measured in \rb \\
posang\_i & real & position angle as measured in \ib \\
posang\_ha & real & position angle as measured in \hb \\
seeing\_flag* & int & flag sources with seeing $<$1.7 \\
photom\_flag$^\dagger$ & int & zero if object detected in all bands \\
class\_flag* & int & combined probabily of stellar object \\
chip\_r\_id* & long & link to \rb chip table \\
chip\_i\_id* & long & link to \ib chip table \\
chip\_ha\_id* & long & link to \hb chip table \\
cx$^\dagger$ & float & Spherical coordinate x=$cos(\alpha)sin(\delta)$\\
cy$^\dagger$ & float & Spherical coordinate y=$sin(\alpha)sin(\delta)$\\
cz$^\dagger$ & float & Spherical coordinate z=$cos(\delta)$ \\
htmid* & long & 20-deep HTM index of this object\\
field\_id* & long & link to field table \\ \hline
\end{tabular}
\caption {Column names, types and description of main photometric tables \textbf{PhotoObj} and \textbf{PhotoObjBest}. Columns marked with an asterisk are indexed. Columns marked with $\dagger$ are not included in the \textbf{PhotoObjBest} table.}
\label{tab:photoobj}
\end{table}

\begin{table}
\centering
\begin{tabular}{@{}rlr@{\ }r@{\ }r@{\ }r@{}} \hline
Index & Meaning & \multicolumn{4}{c}{Probability (\%)} \\
        &                & Star & Extended & Noise & Saturated \\ \hline
-9 & Saturated & 0 & 0 & 5 & 95 \\
-3 & Probable extended & 25 & 70 & 5 & 0 \\
-2 & Probable star & 70 & 25 & 5 & 0 \\
-1 & Star & 90 & 5 & 5 & 0 \\
0 & Noise & 5 & 5 & 90 & 0 \\
1 & Extended & 5 & 90 & 5 & 0\\ \hline
\end{tabular}
\caption{Meaning of classification index and probability assigned to each of them.}
\label{tab:class}
\end{table}

\section{Cross matching with UKIDSS GPS}
\label{sec:science}

In \cite{2005MNRAS.362..753D} we describe the importance of the IPHAS data by themselves and the several papers published to date using just IPHAS data are clearly a demonstration of that.
In this section however, we want to focus in the multiwavelength aspect which the present IPHAS release will facilitate. We provide an example of cross matching the sources extracted from the IPHAS catalogue with those from the UKDSS GPS. Near infrared data are required to 
probe the many parts of the galaxy which suffer high extinction and an ESO-public near infrared survey of 
the northern plane, the UKIDSS Galactic Plane Survey, has recently begun.

We have selected a contiguous area of 2 square degrees centred at $\alpha=20^h 48^m, \ \delta=42^\circ 45^\prime$ ($l,b = 83.0, -0.5$) in the Cygnus-X region and extracted all objects from the \textbf{PhotoObjBest} catalogue. 
From the \cite{1998ApJ...500..525S} maps the extinction is $E(B-V)=3$. 
The total number of objects in that area is 235,231. 
In order to assess the photometry we have grouped all the objects by field and used the field overlap algorithm as described in \cite{1994MNRAS.266...65G} in order to offset the photometry to a common reference system. 

For this exercise we have just set one frame as the reference calibrated frame and allowed the others to vary and solved the matrix equation for each band. The resultant magnitude offsets between the different fields is smaller than $\sim$0.03 in the three bands confirming the good photometric calibration.

\begin{figure}
\includegraphics[trim=30 0 40 20,clip=true,width=\hsize]{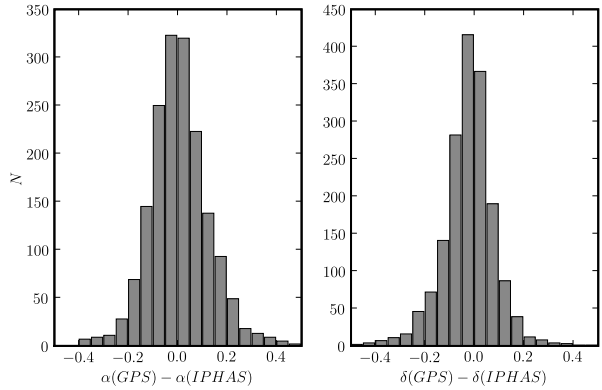}
\caption{Positional offsets in R.A. (left) and Dec (right) between GPS and IPHAS positions (arcsec) in one of the fields included in the area considered in the text.}
\label{fig:offsets}
\end{figure}

For the same area we extract also the near-IR magnitudes from the UKIDSS GPS survey. The GPS is carrying out a complementary survey of the galactic plane in J, H and K. Using \ag access to the WFCAM Science Archive (WSA) we download the near-IR catalogue of objects detected in the area covered by the IPHAS observations. 
We then crossmatch the GPS and IPHAS catalogues selecting the nearest neighbour in a radius of 1.2 arcsec using STILTS\footnote{STILTS is a set of command line tools to perform operations with tables like cross matching (http://www.star.bris.ac.uk/$\sim$mbt/stilts/)}. Figure~\ref{fig:offsets} shows the positional offsets between the IPHAS and GPS catalogues for one field. The offset is effectively zero and the rms is 0.13" and 0.12" in $\Delta\alpha$ and $\Delta\delta$ respectively. This good agreement is not coincidence since both surveys use 2MASS as the reference astrometric system and the WFCAM processing pipeline utilises the same algorithms developed at CASU for the optical processing. It is however worth noting the success of the astrometric pipeline in such different data products.

\begin{figure*}
\includegraphics[trim=20 10 30 10,clip=true,width=0.48\hsize]{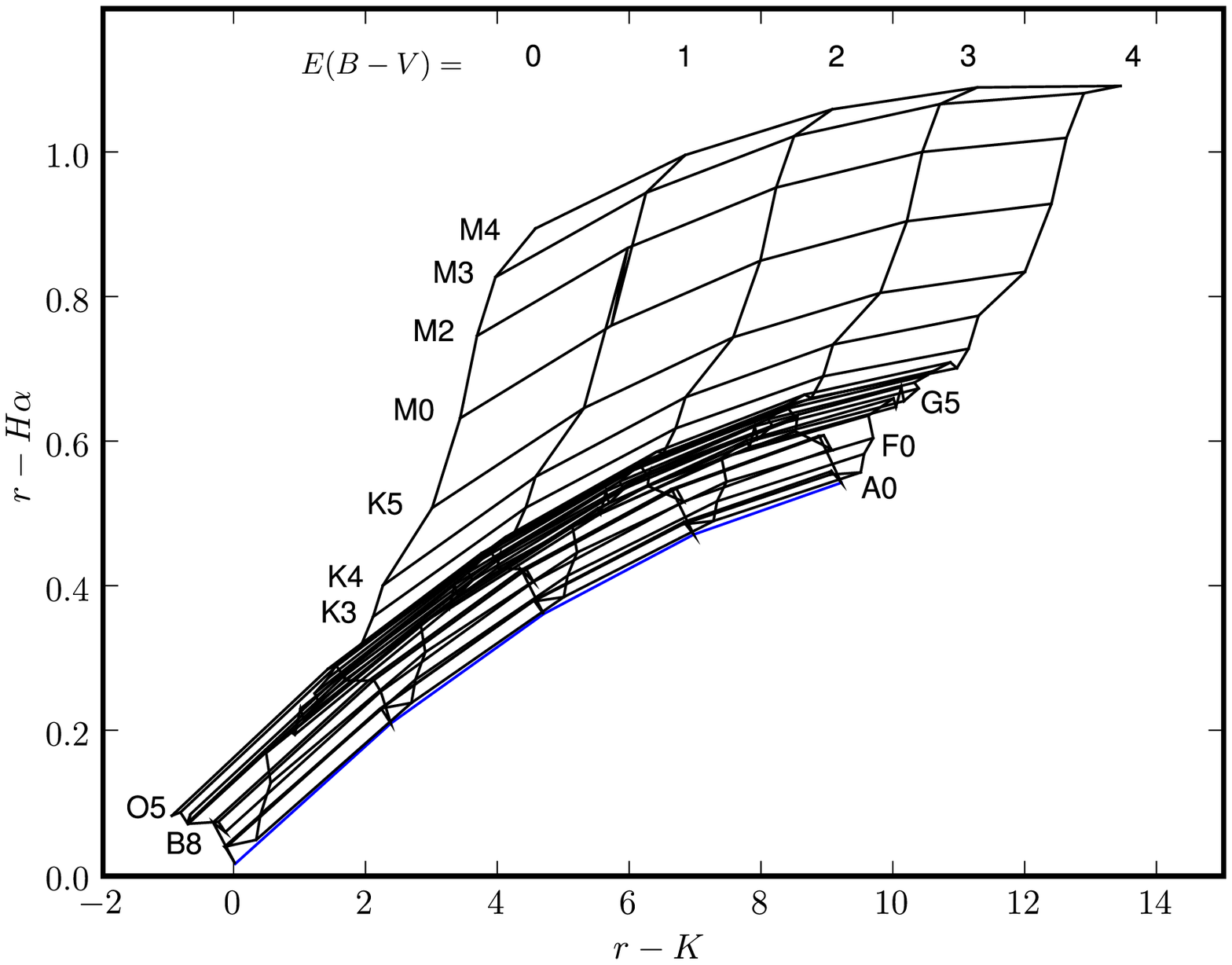}
\includegraphics[trim=20 10 30 10,clip=true,width=0.48\hsize]{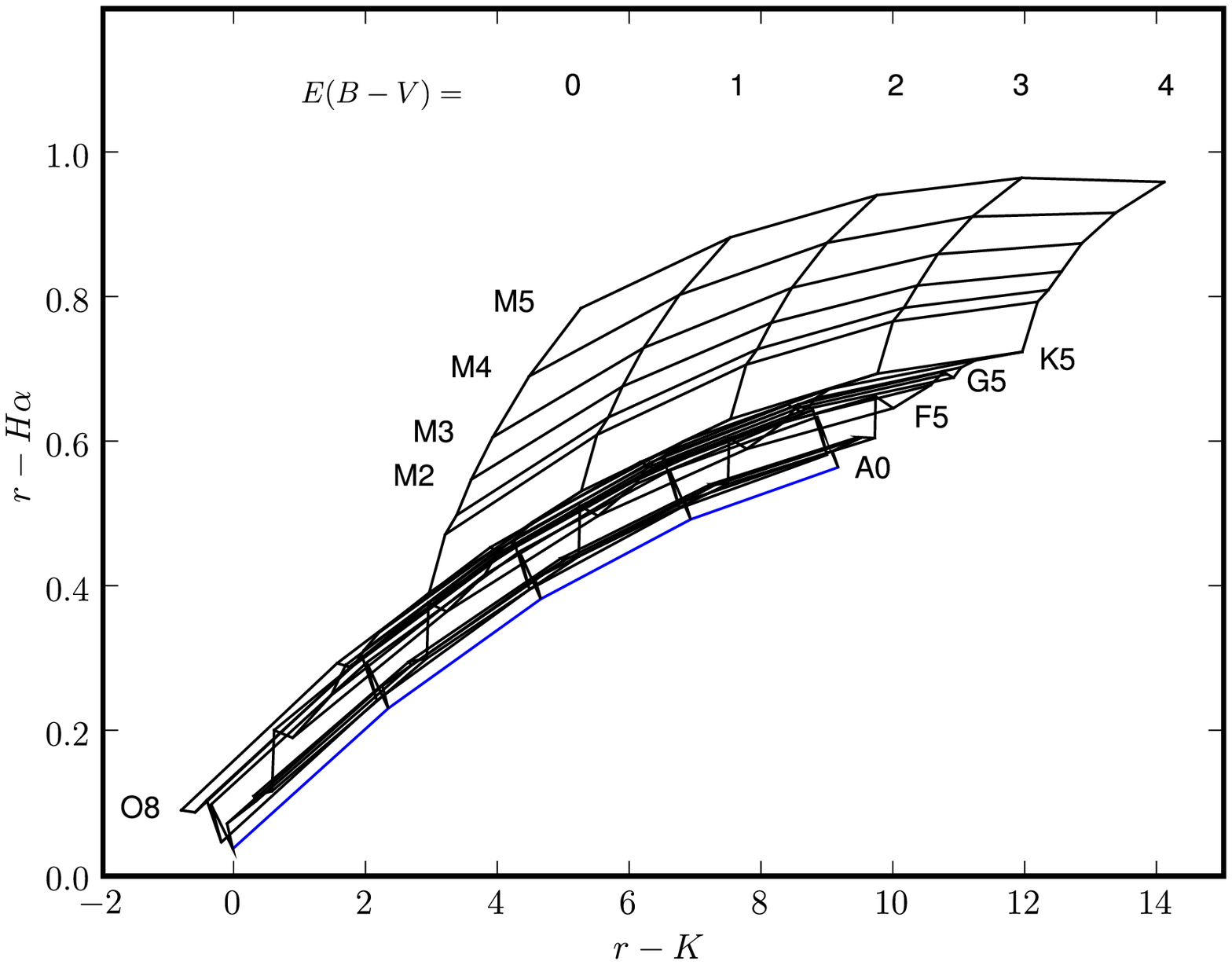}
\caption{Reddening tracks in the $r-H\alpha$ versus $r-K$ plane for main sequence dwarfs (left) and giants (right).}
\label{fig:tracks}
\end{figure*}

\begin{figure*}
\includegraphics[trim=20 10 30 10,clip=true,width=0.48\hsize]{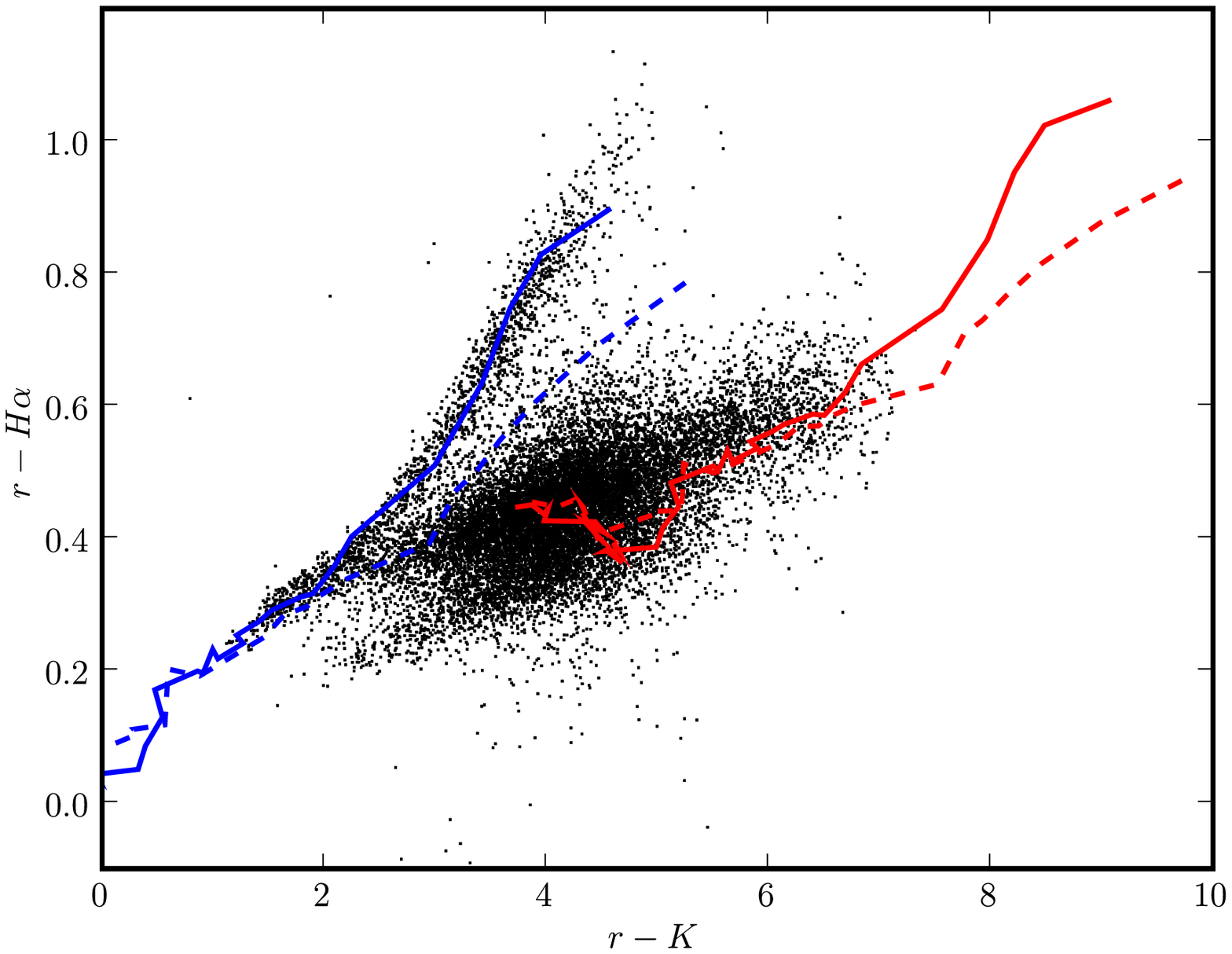}
\includegraphics[trim=20 10 30 10,clip=true,width=0.48\hsize]{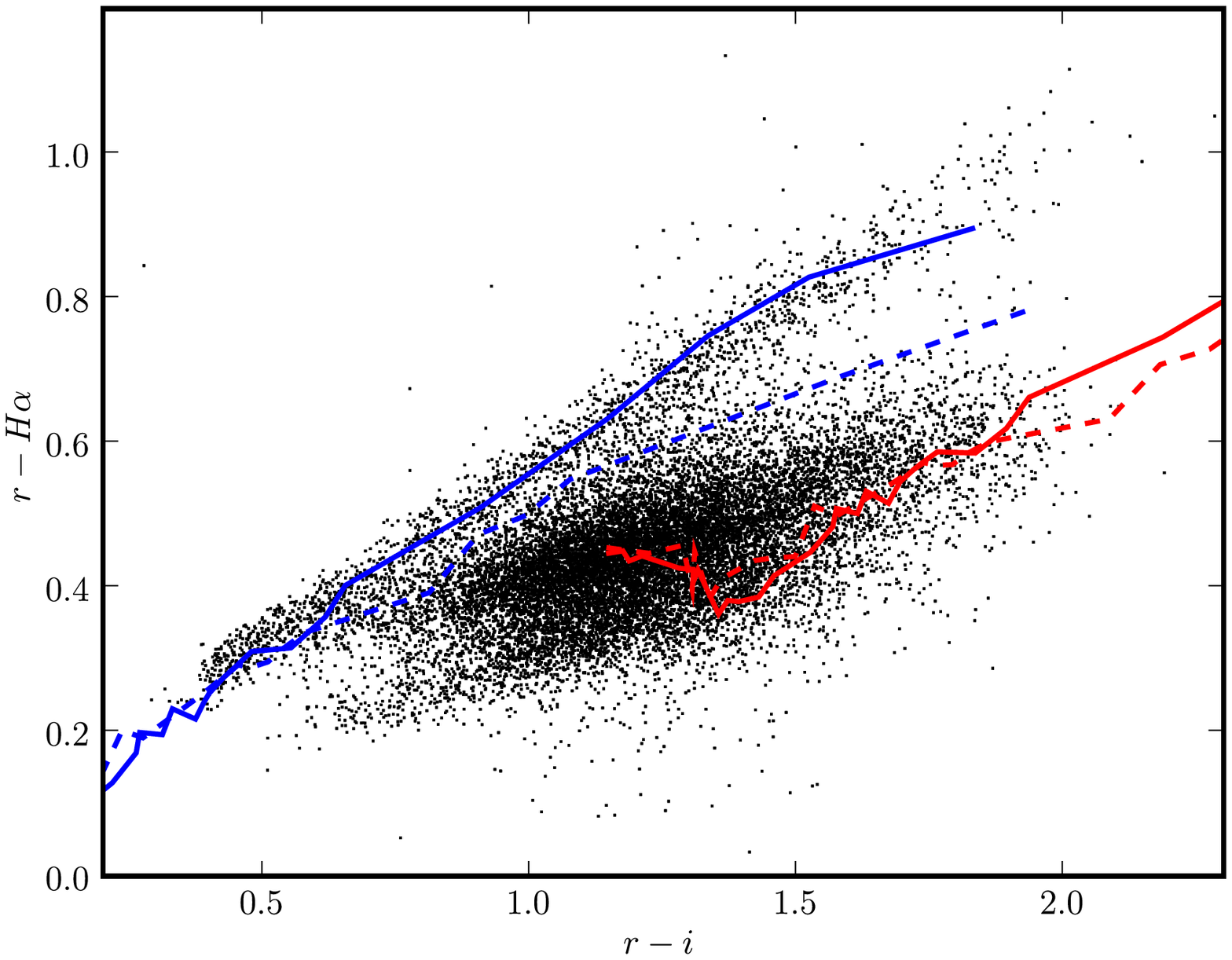}
\caption{Colour colour plots for point like sources detected in IPHAS and GPS with main sequence tracks overlayed (continuous line: main sequence dwarfs; dashed-line: giants) for reddening: $E(B-V)$ = 0 (blue) and 2 (red))
}
\label{fig:ccplots}
\end{figure*}

The importance of the ($r' - H\alpha$, $r' - i'$) plane has been extensively discussed in 
\cite{2005MNRAS.362..753D}. The mean sequence is clearly defined and sources above it are emission line objects (see figure~\ref{fig:ccplots}).
We explore here the ($r'-H\alpha$, $r' - K$) plane provided by the combination of IPHAS and GPS. 

Using the SED model library from \cite{1998PASP..110..863P} and the WFCAM filter profiles we have simulated the colours of normal stars. The detailed process is described in \cite{2005MNRAS.362..753D} and Sale et al. 2008 (in prep) and we list in table~\ref{tab:colours} the synthetic colours for dwarfs and giants only. The tracks are plotted in figure~\ref{fig:tracks} for extinction between zero and four. Figure~\ref{fig:ccplots} shows the ($r' - H\alpha$, $r' - K$) and ($r' - H\alpha$, $r' - i'$) colour-colour diagrams with $E(B-V)=0$ and $E(B-V)=2$ reddening tracks for dwarfs and giants overplotted. We have included in these plots only sources classified as point-like in the
optical and near-IR data (\verb+class_r=-1, -2+, \verb+pStar>0.7+) with magnitude errors smaller than 0.04
in \rb\ and $K$. We have also excluded objects brighter than \rb=13 and $K=12$ to avoid saturation and finally we have removed all sources within 10 arcsec of any WFCAM chip edge (although problematic sources in the edges are anyway flagged by the pipeline classifier).

The main sequence of objects at zero reddening is well defined in the ($r'-H\alpha$, $r' - K$)
plane with a large population of objects up to the M stars. 
These data are for fields in the
Cygnus-X part of the plane where there is a local essentially unreddened
stellar population to a few hundred parsecs, and then a decline in stellar
density and sharp rise in molecular gas and extinction to a distance of
a kpc or so.  Beyond this, the stellar density recovers as distant 
associations are traversed but of course it presents as a much more heavily
reddened population. Furthermore stars later than spectral type mid-K are intrinsically faint and are only detectable at $E(B-V) <~0.8$ in the figure~\ref{fig:ccplots} where our limit is at $r' - K < 7$. 
This explains the lack of K and M dwarfs reddened objects in these diagrams.

Note that no artificial offsets have been introduced to the synthetic tracks to make them fit with the observed colours. This is further confirmation of the good photometric calibration of both the IPHAS and GPS catalogues.

\begin{table*}
\centering
\begin{tabular}{lrrrrrrrrrr} \hline
Spectral & \multicolumn{10}{c}{Reddening} \\
type & \multicolumn{2}{c}{$E(B-V)=0.0$} &  \multicolumn{2}{c}{$E(B-V)=1.0$} &  \multicolumn{2}{c}{$E(B-V)=2.0$} &  \multicolumn{2}{c}{$E(B-V)=3.0$} &  \multicolumn{2}{c}{$E(B-V)=4.0$} \\
 & $(r' - J)$ & $(J-K)$ & $(r' - J)$ & $(J-K)$ & $(r' - J)$ & $(J-K)$ & $(r' - J)$ & $(J-K)$ & $(r' - J)$ & $(J-K)$ \\ \hline
O5 V & -0.955& -0.331 &  1.415&  0.246 &  3.742&  0.820 &  6.028&  1.391  &  8.275 &  1.959 \\
O9 V & -0.826& -0.252 &  1.543&  0.323 &  3.869&  0.895 &  6.154&  1.465 &  8.401 &  2.032\\
B0 V & -0.720& -0.139 &  1.649&  0.437 &  3.976&  1.010 &  6.262&  1.580 &  8.509 &  2.147\\
B1 V & -0.674& -0.107 &  1.690&  0.467 &  4.011&  1.038 &  6.291&  1.607 &  8.533 &  2.172\\
B3 V & -0.706& -0.283 &  1.658&  0.294 &  3.980&  0.868 &  6.262&  1.439 &  8.505 &  2.008\\
B8 V & -0.249&  0.000 &  2.030&  0.531 &  4.423&  1.144 &  6.806&  1.715 &  9.044 &  2.281\\
B9 V & -0.145&  0.003 &  2.108&  0.573 &  4.530&  1.147 &  6.697&  1.712 &  8.933 &  2.278\\
A0 V &  0.005& -0.005 &  2.366&  0.569 &  4.684&  1.140 &  6.961&  1.708 &  9.200 &  2.274\\
A2 V & -0.139& -0.088 &  2.220&  0.489 &  4.536&  1.063 &  6.812&  1.634 &  9.050 &  2.204\\
A3 V & -0.155& -0.131 &  2.202&  0.446 &  4.517&  1.020 &  6.791&  1.591 &  9.028 &  2.160\\
A5 V &  0.324&  0.085 &  2.678&  0.658 &  4.990&  1.228 &  7.261&  1.796 &  9.494 &  2.361\\
A7 V &  0.390&  0.132 &  2.739&  0.705 &  5.047&  1.275 &  7.314&  1.842 &  9.544 &  2.407\\
F0 V &  0.543&  0.159 &  2.888&  0.731 &  5.192&  1.300 &  7.455&  1.867 &  9.683 &  2.432\\
F2 V &  0.475&  0.057 &  2.819&  0.632 &  5.122&  1.205 &  7.384&  1.775 &  9.612 &  2.343\\
F5 V &  0.859&  0.221 &  3.202&  0.793 &  5.503&  1.362 &  7.765&  1.929 &  9.991 &  2.493\\
F6 V &  0.909&  0.218 &  3.249&  0.788 &  5.547&  1.357 &  7.806&  1.922 & 10.029 &  2.486\\
F8 V &  0.995&  0.230 &  3.333&  0.802 &  5.630&  1.372 &  7.887&  1.939 & 10.109 &  2.504\\
G0 V &  1.042&  0.275 &  3.378&  0.847 &  5.674&  1.416 &  7.930&  1.982 & 10.150 &  2.546\\
G2 V &  1.271&  0.329 &  3.606&  0.899 &  5.899&  1.467 &  8.154&  2.031 & 10.373 &  2.594\\
G5 V &  1.210&  0.310 &  3.543&  0.882 &  5.836&  1.451 &  8.089&  2.018 & 10.308 &  2.582\\
G8 V &  1.415&  0.398 &  3.746&  0.968 &  6.035&  1.536 &  8.286&  2.101 & 10.502 &  2.664\\
K0 V &  1.528&  0.499 &  3.853&  1.068 &  6.138&  1.635 &  8.385&  2.198 & 10.598 &  2.760\\
K2 V &  1.790&  0.550 &  4.114&  1.118 &  6.398&  1.683 &  8.643&  2.246 & 10.855 &  2.807\\
K3 V &  1.899&  0.556 &  4.220&  1.125 &  6.501&  1.691 &  8.744&  2.254 & 10.954 &  2.815\\
K4 V &  2.096&  0.648 &  4.411&  1.216 &  6.686&  1.781 &  8.924&  2.344 & 11.128 &  2.904\\
K5 V &  2.245&  0.674 &  4.560&  1.241 &  6.835&  1.806 &  9.074&  2.368 & 11.280 &  2.928\\
M0 V &  2.992&  0.781 &  5.296&  1.348 &  7.561&  1.912 &  9.789&  2.474 & 11.985 &  3.033\\
M2 V &  3.418&  0.807 &  5.961&  1.375 &  7.971&  1.942 & 10.194&  2.506 & 12.387 &  3.067\\
M3 V &  3.953&  0.820 &  6.237&  1.388 &  8.482&  1.953 & 10.695&  2.516 & 12.878 &  3.076\\
M4 V &  4.560&  0.832 &  6.830&  1.399 &  9.062&  1.965 & 11.261&  2.527 & 13.433 &  3.088\\ 
 & & & & & & & & & & \\
O8 III    & -0.813 & -0.076    &  1.550 &  0.496   &  3.870 &  1.065 &  6.149 &  1.631  &  8.390 &  2.195 \\
B1-2 III   & -0.603 & -0.198    &  1.766 &  0.376	  &  4.092 &  0.947 &  6.377 &  1.516  &  8.623 &  2.082\\
B3 III    & -0.427 &  0.018    &  1.935 &  0.591	  &  4.253 &  1.161 &  6.530 &  1.729  &  8.769 &  2.295\\
B9 III    & -0.206 & -0.070    &  2.153 &  0.504	  &  4.469 &  1.076 &  6.744 &  1.644  &  8.982 &  2.210\\
B5 III    & -0.361 & -0.078    &  2.000 &  0.497	  &  4.318 &  1.068 &  6.595 &  1.637  &  8.835 &  2.203\\
A0 III    & -0.030 &  0.020    &  2.326 &  0.593	  &  4.641 &  1.164 &  6.915 &  1.732  &  9.151 &  2.297\\
A3 III    & -0.119 & -0.139    &  2.238 &  0.437	  &  4.553 &  1.010 &  6.828 &  1.581  &  9.065 &  2.150\\
A5 III    &  0.323 &  0.101    &  2.675 &  0.673	  &  4.985 &  1.243 &  7.254 &  1.810  &  9.486 &  2.374\\
A7 III    &  0.285 & -0.003    &  2.638 &  0.572	  &  4.948 &  1.144 &  7.218 &  1.713  &  9.451 &  2.280\\
F0 III    &  0.565 &  0.199    &  2.911 &  0.770	  &  5.216 &  1.339 &  7.481 &  1.905  &  9.709 &  2.468\\
F2 III    &  0.595 &  0.100    &  2.936 &  0.673	  &  5.236 &  1.242 &  7.496 &  1.809  &  9.720 &  2.373\\
F5 III    &  0.876 &  0.291    &  3.214 &  0.861	  &  5.510 &  1.428 &  7.766 &  1.993  &  9.987 &  2.556\\
G0 III    &  1.458 &  0.424    &  3.791 &  0.992	  &  6.083 &  1.559 &  8.337 &  2.123  & 10.555 &  2.684\\
G5 III    &  1.681 &  0.477    &  4.008 &  1.046	  &  6.294 &  1.612 &  8.542 &  2.175  & 10.756 &  2.736\\
G8 III    &  1.727 &  0.516    &  4.053 &  1.086	  &  6.339 &  1.653 &  8.586 &  2.217  & 10.799 &  2.779\\
K0 III    &  1.844 &  0.541    &  4.167 &  1.110	  &  6.449 &  1.676 &  8.694 &  2.239  & 10.904 &  2.800\\
K1 III    &  1.961 &  0.571    &  4.283 &  1.140	  &  6.564 &  1.707 &  8.808 &  2.271  & 11.018 &  2.832\\
K2 III    &  2.188 &  0.671    &  4.506 &  1.239	  &  6.783 &  1.805 &  9.024 &  2.368  & 11.230 &  2.929\\
K5 III    &  2.948 &  0.864    &  5.253 &  1.431	  &  7.519 &  1.995 &  9.748 &  2.557  & 11.945 &  3.116\\
M0 III    &  3.190 &  0.968    &  5.493 &  1.534	  &  7.756 &  2.097 &  9.983 &  2.657  & 12.177 &  3.215\\
M1 III    &  3.364 &  0.987    &  5.664 &  1.553	  &  7.925 &  2.116 & 10.149 &  2.676  & 12.341 &  3.234\\
M2 III    &  3.586 &  1.068    &  5.881 &  1.633	  &  8.135 &  2.196 & 10.354 &  2.757  & 12.541 &  3.315\\
M3 III    &  3.910 &  1.062    &  6.200 &  1.627	  &  8.450 &  2.190 & 10.664 &  2.750  & 12.846 &  3.307\\
M4 III    &  4.464 &  1.086    &  6.744 &  1.652	  &  8.983 &  2.215 & 11.187 &  2.775  & 13.361 &  3.333\\
M5 III    &  5.249 &  1.142    &  7.515 &  1.707	  &  9.742 &  2.270 & 11.934 &  2.830  & 14.098 &  3.387\\
\hline
\end{tabular}
\caption{Synthetic tracks in the ($r'-J$, $J-K$) plane, for main sequence dwarfs and giants, calculated for a range of reddening. A similar table with optical colours only can be found in table 2 of Drew et al. 2005.}
\label{tab:colours}
\end{table*}

\section{User Access}
\label{sec:access}

\begin{figure*}
\centering
\includegraphics[trim=0 0 0 0, clip=true, width=0.8\hsize]{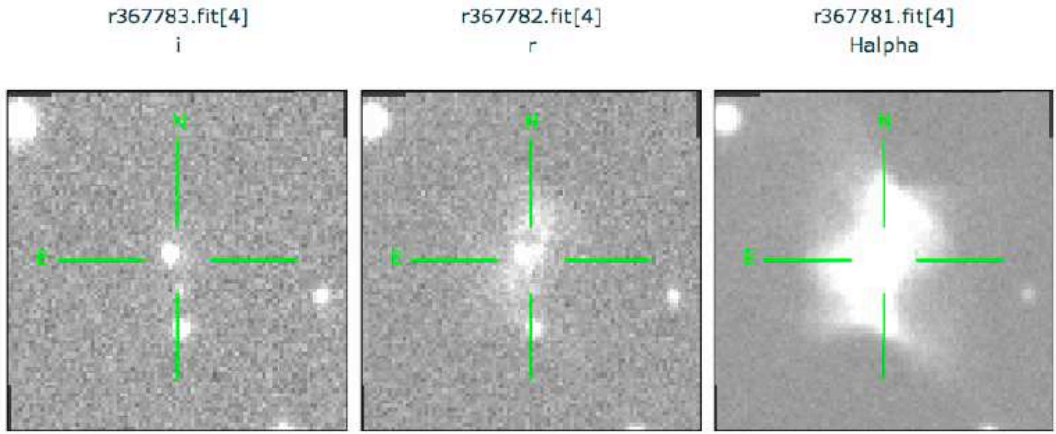}
\caption{Example postage stamps obtained using the finding chart utility. Each stamp is labeled with the run number, CCD number and filter. The object corresponds
to the Principe de Asturias quadrupolar nebula presented in Mampaso et al. (2006).}
\label{fig:postage}
\end{figure*}

User access to the catalogues and images has been provided by the
AstroGrid project.  There are two separate user interfaces to the
IPHAS IDR. Users who only want to extract fluxes or images around a
small position of sky (there is a limit on the radius of 1 degree)
will find the web interface useful\footnote{Available from
  http://www.iphas.org/idr}. Given a position or an object name the
user will obtain a cone search of the \textbf{PhotoObjBest} catalogue
in a variety of output formats (VOTable, Binary VOTable, Comma
Separated Value or HTML). Input lists of objects are also
supported. Together with the IPHAS catalogue we provide access from
the same interface to other useful VO compliant archives like SDSS DR5
\citep{2007ApJS..172..634A}, 2MASS Point Source Catalogue
\citep{2006AJ....131.1163S}, NED and the Carlsberg Meridian
Astrometric Catalogue~\citep{2002A&A...395..347E}.  Postage stamps
generation for a particular input position is also available using the
finding chart utility (see example output in
figure~\ref{fig:postage}).

Users who want more sophisticated queries on flags and algebraic
combinations of parameters (e.g. colours) will find the AstroGrid
\textit{Query Builder} more suited to their needs. The \textit{Query
  Builder} is an application inside the AstroGrid VO Desktop which
allows to submit an arbitrary query to the database system. The
\textit{Query Builder} facilitates the task of building complex
queries written in Astronomical Dataset Query Language
(ADQL\footnote{http://www.ivoa.net/Documents/latest/ADQL.html}; an
international agreed standard based on the Structured Query Language
-- SQL), save the queries and submit them asynchronously to the server
(this facilitates sending long running queries and retrieve the
results later).

\subsection{Image Access}

The Simple Image Access Protocol (SIAP) defines a prototype standard
for retrieving image data from a variety of astronomical image
repositories through a uniform interface\footnote{See
  http://www.ivoa.net/Documents/latest/SIA.html}. The IPHAS images are
available through a SIAP (ivo://uk.ac.cam.ast/IPHAS/images/SIAP)
interface. The result of a query is, given a box centred at
coordinates RA, Dec and a box size in degrees, a table of image CCDs
which overlap the defined box. The table contains links to the
processed image file itself. Alternatively the same list of images can
be obtained with an appropriate SQL query (see below). At the moment
only full image retrieval is supported. Although the graphical
interface does not allow for a list of positions this is possible to
accomplish from the command line interface.

\subsection{Full Catalogue Access}

The ingestion of all the catalogue data into a database provides the
capability of using SQL in order to query the catalogue with any user
defined constraint. The AstroGrid Data Set Access (DSA) provides a
layer on top of any database system which makes it readily VO
compliant and accessible from any VO compliant application. The
examples in appendix A provide ADQL queries that can be used in the Query
Builder to retrieve a variety of data.

\subsection{Command line access}

The AstroGrid Run Time (AR) provides a bridge between the user command
line tools or programs and the VO.  Catalogues can be accessed using
this bridge from programming languages like Python, Perl, C or
Java. There are several reasons to provide a command line interface to
the VO. The more important one is the ability to perform batch queries
for a list of objects and also to implement a call to a VO service
inside a user written script that performs some analysis
afterwards. 

\section{Summary}

We have presented the Initial Data Release of the IPHAS project. The
total release contains photometry in \rb, \ib, and \hb\ for about 200
million objects in the northern galactic plane comprising the largest
photometric sample available imaged in narrowband \hb. We have also
made available a subset of this catalogue, \textbf{PhotoObjBest},
which contains the best observations in terms of final survey quality
as defined in the text and which comprises about 60 per cent of the final
release. The legacy of the survey is increased by the extensive
coverage appearing from other projects spanning the wavelength range
range from X-rays to near-IR and sub-millimeter.

The data have been made available through a VO compliant interface
provided by AstroGrid and accessible to all VO projects.

This initial release will be supplemented with data obtained from
January 2006 onwards at a later date to be announced in the project
web pages.  At the time of writing, data-taking continues.  There will
also be a separate Final Release once the survey is entirely complete
and all finally-accepted data have been corrected onto a
fully-interlaced uniform photometric scale.

More information and examples are available from the IPHAS IDR and
AstroGrid websites at http://www.iphas.org/idr and
http://www.astrogrid.org.

\section*{ACKNOWLEDGMENTS} 

This research has made use of data obtained using, or software provided 
by, the UK's AstroGrid Virtual Observatory Project, which is funded by the 
Science \& Technology Facilities Council and through the EU's 
Framework 6 programme.

This paper makes use of data from the Isaac Newton Telescope, operated
on the island of La Palma by the ING in the Spanish Observatorio del
Roque de los Muchachos of the Instituto de Astrof\'{\i}sica de
Canarias.  IPHAS has been a massive effort during 5 years of
observations and the following are thanked for their contributions at
the telescope: Andrew Cardwell, Jesus Corral-Santana, Pasqual Diago,
James Furness, Antonio Garc\'{\i}a, Wilbert van Ham, Marie Hrudkova,
Avon Huxor, Mansura Jaigirdar, Rob King, Richard Parker, Mark Pringle,
Aaron Robotham, Ernesto Rodriguez-Flores, Miguel Santander, Rachel
Smith and Carolin Winkworth.

PJG, EvdB, GR and LMR are supported by NWO-VIDI grant 639.042.201 to PJG. DS acknowledges a STFC Advanced Fellowship as well as support through
the NASA Guest Observer program.
QP is supported by ANSTO Access to Major Research Facilities Programme (AMRFP).

This work is based in part on data obtained as part of the UKIRT Infrared Deep Sky Survey.

\bibliography{biblio}
\bibliographystyle{mn2e}

\appendix

\section[]{Example Usage Access}

\noindent \textit{Example 1: Return coordinates, magnitudes and magnitude errors of objects in a rectangular area.}

This is the most basic query returning all objects in a specific box with R.A. between 300 and 301 degrees and Dec between 30 and 31 degrees.

\begin{verbatim}
SELECT P.ra, P.dec, 
  P.coremag_r, P.coremag_i, P.coremag_ha,
  P.coremagerr_r, P.coremagerr_i, P.coremagerr_ha
FROM PhotoObjBest as P
WHERE P.ra BETWEEN 300.0 AND 301.0
  AND P.dec BETWEEN 30.0 and 31.0
\end{verbatim}

\noindent \textit{Example 2: Return coordinates and magnitudes of objects in a cone defined by its center at RA=300.0 and Dec=30.0 and radius r=10 arcmin.}

In this case instead of coding the whole trigonometry in the query we use the auxiliary table \textbf{tGetNearbyObj} which returns the object IDs which satisfy a cone search specified by the center coordinates
(ra, dec) and radius (r) in minutes of arc:
 
\begin{verbatim}
SELECT P.ra, P.dec, 
  P.coremag_r, P.coremag_i, P.coremag_ha,
  P.coremagerr_r, P.coremagerr_i, P.coremagerr_ha
FROM PhotoObjBest as P, tGetNearbyObj as G
WHERE G.ra=300.0 AND G.dec=30.0 AND G.r=10.0
  AND P.objID = G.objID
\end{verbatim}

\noindent \textit{Example 3: As previous example but only return objects classified as point-like in r and observed in conditions of seeing best than 1.5 arcsec as measured in the r band.}

In this query we are asking for information about the quality of observations and this is stored in the ChipsAll table. We need therefore to join the results with this table in order to select only observation with good seeing conditions. 

\begin{verbatim}
SELECT P.ra, P.dec, 
  P.coremag_r, P.coremag_i, P.coremag_ha,
  P.coremagerr_r, P.coremagerr_i, P.coremagerr_ha
FROM PhotoObjBest as P, tGetNearbyObj as G, 
  ChipsAll as CR
WHERE G.ra=300.0 AND G.dec=30.0 AND G.r=10.0
  AND CR.seeing<1.5 AND (P.class_r=-1 OR P.class_r=-2)
  AND P.objID = G.objID AND P.chips_r_id=CR.chipID
\end{verbatim}

Adding additional constraints on e.g. object colours, magnitude limits, observation dates are trivially implemented in the same way, e.g.:

\begin{verbatim}
SELECT P.ra, P.dec, 
  P.coremag_r, P.coremag_i, P.coremag_ha,
  P.coremagerr_r, P.coremagerr_i, P.coremagerr_ha
FROM PhotoObjBest as P, tGetNearbyObj as G, 
  ChipsAll as CR
WHERE G._ra=300.0 AND G._dec=30.0 AND G._r=10.0
  AND P.coremag_r-P.coremag_ha>0.8 
  AND P.coremag_r-P.coremag_i>1.5
  AND P.coremagerr_r<0.02
\end{verbatim}

\noindent \textit{Example 4: Select all objects observed in a particular offset field.}

The following example returns all objects detected in a particular observation
from the (not cleaned) PhotoObj catalogue.

\begin{verbatim}
SELECT * FROM PhotoObj as P
  JOIN FieldsAll as F ON (F.fieldID = P.field_id) 
WHERE F.fieldno=7012 AND F.onoff=0
\end{verbatim}

\noindent \textit{Example 5: Return the number and position of fields observed in the best photometric conditions.}

Section~\ref{sec:catalogues} explains the constrains used to define the catalogue of best observations. The following query returns the number, position and photometric characteristics of the fields
which satisfy those conditions.

\begin{verbatim}
SELECT 
    CC.night, F.fieldID, F.fieldno, F.onoff, 
    F.ra, F.dec, F.glon, F.glat, 
    T.magzpt_mean, T.magzrr_mean, 
    CC.seeing, CC.elliptic 
FROM
   ( SELECT
       C.night, 
       sum(C.magzpt)/count(C.magzpt) as magzpt_mean, 
       sum(C.magzrr)/count(C.magzrr) as magzrr_mean 
    FROM FieldsAll F, ChipsAll C 
    WHERE
       C.field_id = F.fieldID AND C.band='r' 
       AND C.chipno=4
   GROUP BY night) AS T, 
   ChipsAll CC, FieldsAll F
WHERE 
   T.night = CC.night AND CC.field_id = F.fieldID 
   AND CC.chipno=4 AND CC.band='r' AND CC.seeing<2 
   AND CC.elliptic<0.2 AND T.magzpt_mean>24.2 
   AND T.magzrr_mean<0.05
ORDER BY night
\end{verbatim}

The IDR web site (http://idr.iphas.org) provides further information as well as examples on
accessing the catalogue from the command line.

\label{lastpage}

\bsp

\end{document}